\documentclass[12pt,epsfig,amsfonts]{article}
\usepackage{graphicx}
\usepackage{cite}
\usepackage{color}
\def\be{\begin{equation}}
\def\ee{\end{equation}}
\def\bea{\begin{eqnarray}}
\def\eea{\end{eqnarray}}

\begin{document}

\newcommand{\dlt}{\bigtriangleup}
\newcommand{\beq}{\begin{equation}}
\newcommand{\eeq}[1]{\label{#1} \end{equation}}
\newcommand{\insertplot}[1]{\centerline{\psfig{figure={#1},width=13.0cm}}}
\newcommand{\insertplotsshort}[1]
{\centerline{\psfig{figure={#1},height=5.0cm}}}
\newcommand{\insertplotll}[1]{\centerline{\psfig{figure={#1},height=11.0cm}}}
\newcommand{\insertplotlll}[1]{\centerline{\psfig{figure={#1},height=18.0cm}}}
\newcommand{\Pom}{\mbox{{\it I$\!$P}}}

\parskip=0.3cm


\vskip 0.5cm \centerline{\Large \bf Forward Physics at the LHC;
Elastic Scattering} \vskip 0.2cm  \vskip 0.3cm \centerline{R.
Fiore$^{a\dagger}$, L. Jenkovszky$^{b,\star}$, R. Orava
$^{c,\bullet},$ E. Predazzi $^{d, \diamond},$ A. Prokudin $^{d, \ddagger},$ and O. Selyugin $^{f,\circ}$}

\vskip 0.2cm

\centerline{$^{a}$ \sl  Dipartimento di Fisica, Universit\'{a}
della Calabria} \centerline{\sl Instituto Nazionale di Fisica
Nucleare, Gruppo collegato di Cosenza} \centerline{\sl I-87036
Arcavata di Rende, Cosenza, Italy} \vskip 0.1cm \centerline{$^{b}$
\sl Bogolyubov Institute for Theoretical Physics} \centerline{\sl
Academy of Science of Ukraine} \centerline{\sl Kiev-143, 03680
Ukraine and} \centerline{\sl RMKI, KFKI, POB 49, Budapest 114,
H-1525 Hungary} \vskip 0.1cm \centerline{$^{c}$ \sl Helsinki
Institute of Phyics and University of Helsinki,} \centerline{\sl
PL64, FI-00014 University of Helsinki, Finland}

\vskip 0.1cm
 \centerline{$^{d}$
\sl  Dipartimento di Fisica Teorica, Universit\`a di Torino,}
\centerline{\sl Instituto Nazionale di Fisica Nucleare, Sezione di
Torino} \centerline{\sl Via P. Giuria 1, I-10125 Torino, Italy
}  \vskip 0.1cm
\centerline {\sl $^{f}$ BLTP, Joint Institute for Nuclear
Research,} \centerline {\sl  141980 Dubna, Moscow region, Russia}

\vskip 0.2cm

\begin{abstract}

 The following effects in
 the nearly forward (``soft") region of the LHC are proposed to be investigated:

$\bullet$ At small $|t|$ the fine structure of the cone (Pomeron)
should be scrutinized: a) a break of the cone near $t\approx -
0.1$ ~ GeV$^2,$ due to the two-pion threshold, and required by
t-channel unitarity, and b) possible small-period
oscillations between $t=0$ and the dip region.

$\bullet$ In measuring the elastic $pp$ scattering and total $pp$
cross section at the LHC, the experimentalists are urged to treat
the total cross section $\sigma_t,$ the ratio $\rho$ of real to
imaginary part of the forward scattering amplitude, the forward
slope $B$ and the luminosity ${\cal L}$ as free parameters, and to
publish model-independent results on ${dN/{dt}}.$

$\bullet$ Of extreme interest are the details of the expected
diffraction minimum in the differential cross section. Its
position, expected in the interval $0.4<-t<1$ GeV$^2$ at the level
of about $10^{-2} \ {\rm mb} \cdot$ GeV$^{-2}\div 10^{-1} \ {\rm
mb}\cdot$ GeV$^{-2}$, cannot be predicted unambiguously, and its
depth, i.e. the ratio of $d\sigma/dt$ at the minimum to that at
the subsequent maximum (about $-t=5 \ $GeV$^2$, which is about $5$) is
of great importance.

$\bullet$ The expected slow-down with increasing $|t|$ of the
shrinkage of the second cone (beyond the dip-bump), together with
the transition from an exponential to a power decrease in $-t$,
will be indicative of the transition from ``soft" to ``hard"
physics. Explicit models are proposed to help in quantifying this
transition.

$\bullet$ In a number of papers a limiting behavior, or
saturation of the black disc limit (BDL), was predicted. This
controversial phenomenon shows that the BDL may not be the
ultimate limit, instead a transition from shadow to antishadow
scattering may by typical of the LHC energy scale.

\end{abstract}

\vskip 0.1cm


$
\begin{array}{ll}
^{\dagger}\mbox{{\it e-mail address:}} &
   \mbox{fiore@cs.infn.it} \\
^{\star}\mbox{{\it e-mail address:}} & \mbox{jenk@bitp.kiev.ua} \\

^{\bullet}\mbox{{\it e-mail address:}} & \mbox{rorava@cc.helsinki.fi} \\
^{\diamond}\mbox{{\it e-mail address:}} &
   \mbox{predazzi@to.infn.it} \\
^{\ddagger}\mbox{{\it e-mail address:}} & \mbox{prokudin@to.infn.it} \\
^{\circ}\mbox{{\it e-mail address:}} &
\mbox{selugin@theor.jinr.ru}
\end{array}
$

\newpage
\centerline{\bf CONTENTS} \vspace{-0.2 cm}

\contentsline{section}{\numberline {1.} {Introduction}}{4}
\vspace{-0.2 cm}
 \contentsline{section}{\numberline {2.} {\bf Measuring Strategy at the LHC}}{7}
\contentsline{subsection} {\numberline {2.1.} Elastic proton signature at the LHC}{7}
\contentsline{subsection} {\numberline {2.2.} Extrapolation to the optical point}{10}
\contentsline{subsection}{\numberline {2.3.} Measuring $\rho$}{10}
\contentsline{subsection}{\numberline {2.4.} Elastic scattering scenarios}{11}
\contentsline{subsection}{\numberline {2.5.} Total $pp$ cross section measurement strategy}{11}
\contentsline{subsection}{\numberline {2.6.} $\sigma_{tot}$ measurement}{14}
 \contentsline{subsection}{\numberline {2.7.} Luminosity measurement and monitoring}{14}
\vspace{-0.2cm} \contentsline{section}{\numberline {3.} Forward
physics at the LHC}{15} \vspace{-0.1cm}
\contentsline{section}{\numberline {4.} The forward cone}{19}
 \contentsline{subsection}{\numberline {4.1} The "break" at very small $|t|$}{19}
 \contentsline{subsection}{\numberline {4.12} Oscillations}{20}
\contentsline{section}{\numberline {5.} The Dip-Bump Region,
$t\sim -1\; {\rm GeV}^2$}{23}
 \contentsline{subsection}{\numberline {5.1.} A simple model for the diffraction pattern}{24}
 \contentsline{subsection}{\numberline {5.2.} The ``Protvino'' model}{26}
 \contentsline{subsection}{\numberline {5.3.} The ``Connecticut'' model}{28}
 \contentsline{subsection}{\numberline {5.4.} The ``Dubna Dynamical Model''}{30}
\vspace{-0.2cm}
 \contentsline{section}{\numberline {6.} Intermediate and Large $|t|$}{34}
\vspace{-0.2cm} \contentsline{section}{\numberline {7.} Black Disc
Limit at the LHC?}{36} \contentsline{subsection}{\numberline
{7.1.} Definitions}{36} \contentsline{subsection}{\numberline
{7.2.} The ``Born term''}{38}
\contentsline{subsection}{\numberline {7.3.} Impact parameter
representation, the black disc limit and unitarity}{40}
\contentsline{subsection}{\numberline {7.4.} Saturation at the LHC
(in the DD model)}{42} \vspace{-0.2cm}
\contentsline{section}{\numberline {8.} Summary and Outlooks}{43}
\vspace{-0.2cm}
\contentsline{section}{\numberline{}References}{44}

\newpage
\section{Introduction} \label{s1}

In this paper, some crucial issues that may be useful in preparing
the experiments at the LHC are discussed and clarified.

Measurement of the total proton-proton cross section will be one
of the first priorities at the LHC. The importance of these
measurements is two-fold. First, they are mandatory to fix the
normalization of all subsequent measurements. Furthermore, the
value of the total cross section will drastically narrow the range
 of the existing models with predictions for the total
cross section ranging as  $\sigma_{tot}(14 \;{\rm TeV})=(125\pm
25) \ $mb \cite{Landshofftot_a, Landshofftot_b} or even
more{\footnote{It would be instructive and amusing to make a
comparative compilation of earlier predictions already checked and
thus confirmed or rejected by the existing data!}. The knowledge
of the total cross section will help in selecting a class of
models of diffraction, based on the dominance of multi-gluon, or
Pomeron ({\mbox{{\it I$\!$P}}}) exchange.

Diffractive events, for example, diffractive Higgs production, are
widely believed to produce the cleanest signal of possible new
phenomena \cite{Helsinki_a, Helsinki_b, Helsinki_c, Helsinki_d,
Albrow}.

Quantum ChromoDynamics (QCD), complemented with the Regge pole
theory, and the unitarity condition superimposed, form the
theoretical basis of the strong interaction. Both QCD and the
Regge pole theory need experimental verification to clarify the
role of higher QCD corrections, on the one hand, and to restrict the
existing flexibility in the Regge pole models, on the other hand.
 The scattering amplitude at the LHC can safely be
parameterized by the dominating Pomeron exchange, appended by a
possible tiny Odderon contribution, the contribution from the
secondary Reggeons at LHC being presumably negligible. A
comprehensive introduction to high-energy diffraction can be found
in Ref. \cite{Bar_Pre}.

LHC will be the first accelerator where the relative contribution
from secondary (sub-leading) trajectories (R) will be negligible,
i.e. smaller than the experimental errors. The ratio
$R/{\mbox{{\it I$\!$P}}}$, apart from kinematics, depends
essentially on the difference of the relevant intercepts, however
the above statement holds even for the most conservative (i.e.
large $R/{\mbox{{\it I$\!$P}}}$) ratio, see e.g.
Ref.~\cite{Des_Ilyin}), decreasing with $|t|$ since the Pomeron
slope is smaller than that of the sub-leading Reggeons.

Parametrization of the Pomeron is far from being unique. According
to Ref.~\cite{Jenk}, there is only one Pomeron in the nature
(although its form is not necessarily simple, see \cite{Pomerons1,
Pomerons2, Pomerons3, Rivista, Lia, Kopeliovich}). The data on
deep inelastic scattering from HERA provoked discussions on the
existence of an alternative, ``hard" or ``QCD Pomeron"
\cite{BFKL_a, BFKL_b, BFKL_c, BFKL_d}, needed for the confirmation
of both perturbative QCD calculations and of the ``hard" and
``semi-hard" diffractive physics. It should be remembered that the
properties (parameters, etc.) of the Pomeron in hadronic
collisions (ISR, SPS, Tevatron and LHC) can be determined with a
precision and reliability much higher than that in $ep$
collisions.

The interface and/or transition between soft and hard dynamics is
a key issue of the strong interaction theory. In elastic
scattering at the LHC it is expected to occur in a smooth way,
within the reach of the forthcoming LHC experiments. Roughly
speaking, this region will be characterized by a transition from
an exponential in $t$, through an exponential in $\sqrt{|t|}$
("Orear regime"), fall-off of the differential cross section to a
power behavior, manifesting hard scattering between point-like
constituents of the nucleons. The transition region is not so
simple because of the different unitarization and rescattering
procedures used in the models. Moreover, non-linear trajectories
(that mimic hard scattering), non-perturbative contributions and
higher order perturbative QCD corrections complicate the issue. In
particular, quark model and QCD calculations, apart from a power
behavior in $t,$  indicate \cite{Donnachie} the onset of an
$s-$independent regime in the differential cross section typical
of  the transition from soft to hard collisions. The complexity of
this transition is connected with the deconfinement of quarks and
gluons in nucleons. We argue in this paper that this transition is
expected in the region $(5<-t<15) \ $GeV$^2$, well within reach of
the LHC measurements.

We propose to investigate LHC effects that can be predicted
qualitatively and which can be measured to give definite
quantitative answers concerning the nature of the strong
interaction at large distances. The LHC will rule out many of the
existing model predictions and thus narrow the class of viable
theoretical approaches.

In Secs. 2 and 3 the essential features of the experimental
program concerning forward physics at the LHC are described.

In Sec. 4 we discuss two types of interesting irregularities
 observed in $t$ within the exponential cone. One is the so-called
 ``break"\footnote {Actually,
the ``break" is an approximation to a smooth curvature.} of the cone
(change of its slope) near $-t=0.1$ GeV$^2$ . The
second one concerns the possibility of tiny oscillations
superimposed over the cone.

In Sec. 5 the ``dip-bump'' structure is analyzed. This is a
sensitive region that will help to understand the nature of the
high-energy diffraction and, eventually, to reveal the Odderon,
whose role (see Ref.~\cite{Nicolescu} and references therein) is
often exaggerated, dramatized and confused. It is maintained (see
e.g. Refs.~\cite{Jenk,Rivista}) that the Odderon should exist
simply because nothing forbids its existence. It is not known how
large the Odderon is or how its contribution varies with $s$ and
$t$. Specific theorems and predictions concerning the Odderon can
be found in
Refs.~\cite{Shelk_a,Shelk_b,Shelk_c,Covolan_a,Covolan_b,LN,odd1,odd2,odd3,odd4,Contogouris}.
 Until now it has
been observed directly in a single experiment
\cite{dip_exp_a,dip_exp_b} only, and therefore needs to be
confirmed experimentally.

\begin{figure}[tbh!]
\begin{center}
\includegraphics[width=0.5\textwidth]{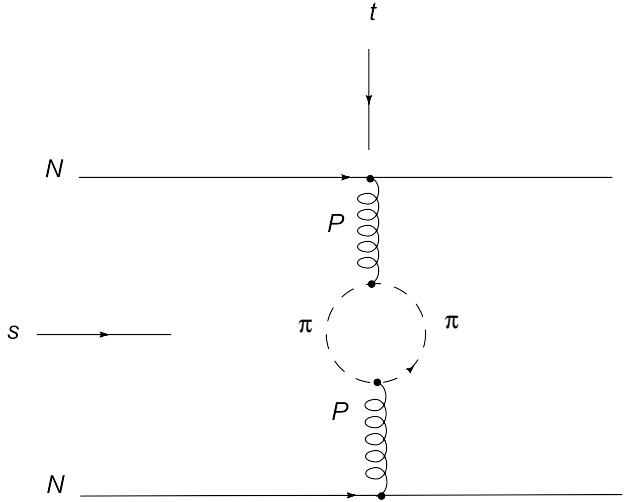}
\end{center}
\caption{Two-pion loop, required by $t-$ channel unitarity, in the
pomeron exchange.} \label{two-pion}
\end{figure}

The region beyond the dip-bump structure, considered in Sec. 6,
along the expected second cone, $-t>1$ GeV$^2$, may be indicative
of the transition from soft to hard physics. It manifests
transition from an exponential to a power decrease in $t$, with a
possible slow-down of the energy dependence \cite{Donnachie}.
Details of this effect will measure the nonlinearity of the
Pomeron trajectory at large enough $|t|$. One should note that the
analytic $S-$matrix theory, perturbative QCD
and the data require that Regge trajectories be nonlinear, complex
functions \cite{nonlin1, nonlin2} (for more details see
Refs.~\cite{Jenk, VJS1}).

At the LHC energies a new phenomenon, namely the  onset of the
Black Disk Limit (BDL) may come into play, changing the
$t$-dependence of the slope $B(s,t).$ This phenomenon is discussed
in details in
 Sec. 7.

The Pomeron trajectory has threshold singularities, the lowest one
being due to the two-pion exchange, required by the $t-$channel
unitarity \cite{nonlin1, nonlin2, Gribov_a, Gribov_b, Gribov_c,
Gribov_d}, as shown in Fig.~\ref{two-pion}. There is a constrain
\cite{nonlin1,Barut_a,Barut_b,Barut_c,Barut_d}, following from the
$t-$ channel unitarity, by which
\begin{eqnarray}
\Im\alpha(t)\sim (t-t_0)^{\Re\alpha(t_0)+1/2}, \ \ t\rightarrow
t_0,
\end{eqnarray}
where $t_0$ is the lightest threshold in the given channel. For
the Pomeron trajectory the lightest threshold is $t_0=4m^2_{\pi}$,
as shown in Fig.~\ref{two-pion}, and the trajectory near the
threshold can be approximated by a square root:
\begin{eqnarray}
\alpha(t)\sim\sqrt{4m^2_{\pi}-t}. \label{sqrt}
\end{eqnarray}
The observed nearly linear behaviour of the trajectory is promoted
by higher, additive thresholds (see Ref.~\cite{traject_a,
traject_b} and references therein).

This threshold singularity appears in different forms in various
models, see Sec. 4. It is also noted that, irrespective of the
specific form of the trajectory or scattering amplitude, the
presence of the above-mentioned threshold singularity in $t$
results in an exponential asymptotic decrease of the impact
parameter amplitude, with important physical consequences (the
so-called nucleon atmosphere, or its clouding).

The important role of nonlinear trajectories and their observable
consequences were first studied in Refs. \cite{nonlin2, slope1,
slope2, slope3, slope4}. Independently, they were also developed
by the Kiev group (see Ref.~\cite{DAMA_a,DAMA_b,DAMA_c} and
references therein), and more recently in a series of papers
(Ref.~\cite{Burak} and references therein).

Asymptotically, the trajectories are logarithmic. This asymptotic
behaviour follows from the compatibility of the Regge behavior
with the quark counting rules \cite{scaling} as well as from the
solution of the BFKL equation \cite{BFKL_a,BFKL_b,BFKL_c,BFKL_d}.
A simple parametrization combining the linear behaviour at small
$|t|$ with its logarithmic asymptotic is \cite{Jenk, VJS1,JenkChi}
\begin{eqnarray}
\alpha(t)=\alpha_0-\gamma\ln(1-\beta_1 t). \label{log}
\end{eqnarray}
Such a trajectory, being nearly linear at small $|t|$, reproduces
the forward cone of the differential cross section, while its
logarithmic asymptotic provides for the wide-angle scaling
behavior \cite{scaling, JenkChi, Fioreetal}. Eqs. (\ref{sqrt})
and (\ref{log}) can be combined in the form \cite{Jenk,Rivista}
\begin{eqnarray}
\alpha(t)=\alpha_0-\gamma\ln(1+\beta_2\sqrt{t_0-t}),
\label{combined}
\end{eqnarray}
where $\beta_1,\ \beta_2$ and $\gamma$ in Eqs. (\ref{log}) and
(\ref{combined}) are parameters whose numerical values can be
associated with their physical meaning (see \cite{JenkNP}).

In a limited range, especially at small and intermediate values of
their argument, linear trajectories may be a reasonable
approximation to their otherwise complex form.

\section{Measurement Strategy at the LHC}

The elastic proton-proton interactions are measured, and
triggered, by the leading proton detectors,
 Roman Pots, placed symmetrically on both sides of the CMS experiment
 at $\pm 147$ and $\pm 220$ meters from the Interaction Point (IP5)
 \footnote{
 The detector locations at  $\pm 145$ and $\pm 149$ meters from IP5 are here
 referred as the ``$\pm 147$'' meter location and the ones at  $\pm 218$
 and $\pm  222$ meters from IP5 as the ``$\pm 220$'' meter location.
 Initially the  $\pm 220$ m location will be instrumented.
 }.
To measure protons at small scattering angles, the detectors must
be moved close to the primary LHC beam in  vertical direction
\footnote{ The closest approach to the beam is of the order of a
few mm's ($10\sigma  + 0.5$ mm) and depends on the measurement
location and used beam optics scheme (see  Ref. \cite{2.1} ). }.
The ``nominal'' TOTEM beam optics set-up has high  $\beta^{*}$ (
$\beta^{*} \sim   1540 \ $ m) and no crossing angle at the IP for
optimizing the acceptance and accuracy at small values of the
four-momentum transfer squared down
 to $ -t_{min}  \approx  2 \cdot   10^{-3} \ $ GeV$^2$.

The $ -t$ distribution of the scattered protons, $ dN_{el}/dt$, is
extrapolated to $-t = 0$, where it is related to the total
proton-proton cross section by the Optical theorem. With the
special optics of  $\beta^{*}= 90 \ $m,
  compatible with the LHC injection optics,
 a first quick measurement of the elastic cross section $d \sigma_{el}/dt$ could be made;
 extrapolation to the Optical point will be made with an accuracy of a few percent.

Runs with different LHC optics set-ups, such as $\beta^{*} = 90 \
$m, the ``injection'' optics ($\beta^{*} = 11 \ $ m),  stage-1
``pilot'' run optics ( $\beta^{*} = 2\ $ m) and the standard LHC
optics ($\beta^*=0.55\ $~m) will allow measurements up to $-t
\approx 10 \div 15 \ $GeV$^2$. Runs with a reduced center-of-mass
energy will allow an analysis of energy dependence, comparisons
with the Tevatron results and a precise measurement of the $\rho$
parameter.

\subsection{Elastic proton signature at the LHC}

Transverse position of an elastically scattered proton with the
momentum loss
 $\xi = \Delta p/p$ at a distance $s$ from the interaction point (IP):
 ( $x(s)$, $y(s)$), is given by the initial coordinates at the IP,
($x^{*}(s=0)$, $ y^{*}(s=0)$), scattering angle, $
\theta^{*}_{x,y}$, the effective length $L^{eff}_{x,y}$,
magnification, $v_{x,y}$, and dispersion, $D$, as (see Ref. \cite{2.1}
and references therein)
\begin{eqnarray}
y(s) &=& \nu_{y}(s) \  y^{*} +L^{eff}_{y}(s) \ \theta^{*}_{y}  \nonumber \\
x(s) &=&\nu_{x}(s) \ x^{*} +L^{eff}_{x}(s) \
\theta^{*}_{x}+\frac{\Delta p}{p} \ D(s)
            \label{2.1}
\end{eqnarray}
By using the transfer matrix, $T(s)$, the initial coordinates of
an elastically scattered proton at the IP can be mapped into a
detector location at point s along the machine. By measuring the
transverse position and scattering angle at point s, the initial
coordinates can be determined. Since angles
 $\theta^{*}_{x,y}$ are very small, they have to be measured by combining
 the symmetrically located detector stations on both sides of the IP:
 a combined measurement, using the left and right arms of the leading proton detectors,
 yields an accurate measurement of the collinear pairs of
  elastically scattered protons (Fig. 2).

From Eq. (\ref{2.1})
\begin{eqnarray}
\theta^{*}_{x} = \frac{x_{R}-x_{L}}{2L^{eff}_{x}}; \ \ \
\theta^{*}_{y} = \frac{y_{R}-y_{L}}{2L^{eff}_{y}}.
            \label{1.13}
\end{eqnarray}

The elastically scattered protons have to be measured close to the
primary LHC beam; down to angles of $5$ to $10 $  $\mu$rad with
respect to the beam direction. For this, special beam optics
conditions with reduced beam divergence at the interaction point
(IP) - and sufficiently large displacement of the scattered
protons at the detector locations (Fig. \ref{fig:lhcacceptance}) -
are required. Thorough studies of optimal LHC beam conditions have
led to the nominal ``TOTEM'' beam optics with  $\beta^{*} \sim
1540 \ $m \footnote{
   There is significant uncertainty in determining the exact value of $\beta^{*}$,
   especially at large values of $\beta^{*}$, and the number given should be taken
   as an approximate figure, only.
 }
 for the measurement of elastic scattering and soft diffraction Ref. \cite{2.1}.

\begin{figure}[tbh!]
\begin{center}
\includegraphics[width=0.9\textwidth]{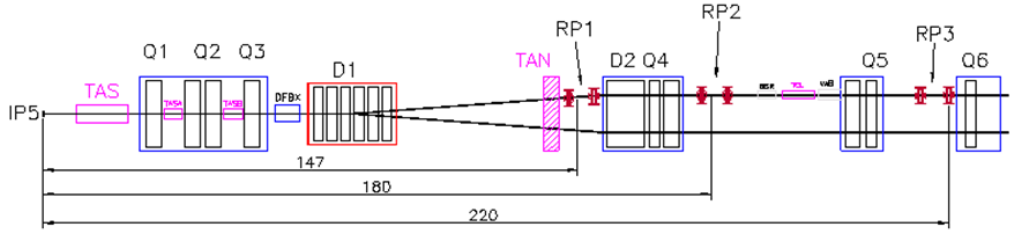}
\end{center}
\caption{The
TOTEM lay-out of leading proton detectors (Roman Pots). The
detector locations at
  $\pm 147 $ m (RP1)  and at $\pm 220 $ m (RP3) are shown  \cite{2.1}.
 The location of RP2 is not part of the present design.}
\label{fig:lhcacceptance}
\end{figure}

 In Fig. 3, the acceptance of elastic protons is shown as a function of
 $-t$ for three different run scenarios.
An acceptance down to $ \sim 5 \cdot 10^{-3} \ $ GeV$^2$ is
achieved with the nominal LHC beam emittance; with improved beam
emittance an acceptance down to $-t_{min} = 2 \cdot 10^{-3} \
$GeV$^2$ is obtained \cite{2.2}.
 Contrary to the nominal TOTEM optics, the  $\beta^{*}= 90 \ $m optics uses the standard
 LHC injection optics and could be realised during the run-in phase of the machine.
 The acceptance in $-t$ reaches $-t_{min} \approx 3.0 \cdot  10^{-2} \ $GeV$^{2}$.

\begin{figure}[tbh!]
\begin{center}
\includegraphics[width=0.9\textwidth]{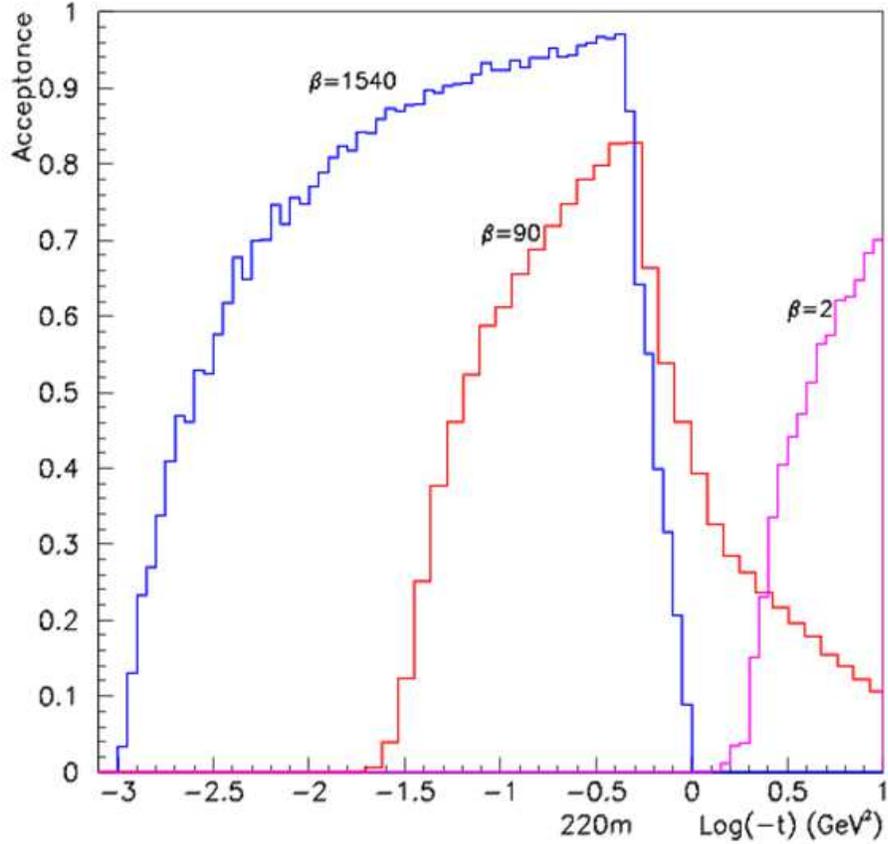}
\end{center}


\caption{$-t$ acceptance of elastic protons detected at $\pm 220 $
m from the IP5 for the 'nominal' ($\beta^* =1540 $ m) and for two
'custom' run options ($\beta^{*}\simeq 90 \ $m.) \cite{2.2}.}

\end{figure}

Due to the geometric constraints imposed by the LHC vacuum pipe
and beam screen, elastic protons with $ -t$ values in excess of $
0.5 \ $GeV$^2$ cannot be measured with the nominal high-$\beta^{*}
$ optics conditions. Also, since at high $-t$ values
$d\sigma_{el}/dt  \propto  1/t^8 $, higher luminosities would be
desired.
 The ``standard'' LHC optics scenario includes the ``injection'' optics with
 $\beta^{*}=11 \ $m and a ``pilot'' stage optics
 with  $\beta^{*}= 2\ $m that allows elastic protons with higher $-t$
  values to be accessed (Figure 3).
  With the $2 \ $m optics, a reasonable event statistics up to
  $-t = 10 \ $GeV$^2$ ($15 \ $GeV$^2$)
  and resolution  $\sigma(|t|)/|t| \approx   2\% $ is achieved.

With the high-$\beta^{*}$ TOTEM optics, so-called {\it
parallel-to-point} focussing conditions are achieved (at  $\pm 220
\ $m in both horizontal and vertical planes) and the elastic
proton measurement becomes independent of the location of the
primary interaction vertex. This will reduce systematic
uncertainty in measuring the four-momentum-transfer squared of the
scattered proton.

The angular beam divergence dominates the uncertainty in measuring
-t, and together with the detector resolution, accounts for
basically all of the uncertainty in $ -t$. For small values of
$-t$ ($-t = 0.01 \ $GeV$^2$)
 the error in $ -t$, $\Delta t/t$, is $\sim 3 \% $
  in case the forward-backward pair of leading proton measurement stations
  is used and $\sim 5 \% $
   if only one of the two leading proton spectrometry ``arms'' is used.

With lower center-of-mass energies, the acceptance in $-t$
improves and, at the energies of  $\sqrt{s} = 2 \ $TeV, $-t$
values of one order of magnitude lower than at the nominal LHC
energy are reached. This would allow comparisons with the Tevatron
results and the Coulomb-hadronic interference region to be probed.

\subsection{Extrapolation to the optical point}

For the total pp cross section measurement, extrapolation of the
elastic scattering $t$ distribution, $dN_{el}/dt$, to $t = 0$ is
required. The relative statistical uncertainty of the
extrapolation is estimated to be
 $ ~ 0.1 \% $
 based on short periods of data taking with the nominal TOTEM optics and luminosity of
 $10^{28} \ $cm$^{-2}$$s^{-1}$   \cite{2.1}.

A thorough study of the systematical effects in the extrapolation
process was carried out  \cite{2.1,2.2} and concluded that a
precision better than $ 0.5 \% $
 should be achieved in extrapolating the elastic cross section to the optical point\footnote{
 For a discussion on model dependences in extrapolation to the Optical point see Ref.
 \cite{2.3}.}.

In case Coulomb scattering is not accounted for in the
extrapolation process, a shift of
 $ (d \sigma_{el}/dt)|_{t=0} \ $of $ (1\div 2) \% $
could occur. Due to Coulomb scattering contribution, the slope $B$
in  $d \sigma_{el}/dt \sim \exp(Bt) $ does not stay constant at
low $-t$ as assumed in a usual extrapolation. The uncertainty due
to this effect might represent an uncertainty of several parts
 in $10^{-3}$ to the extrapolation.

\subsection{Measuring $\rho$ }

For measuring the  $\rho$-parameter,
\begin{equation}
\rho={\Re A(s,t=0)\over{\Im {A(s,t=0)}}},
\end{equation}
one fits the elastic differential cross section:
\begin{equation}
\frac{1}{L}\frac{dN_{el}}{dt}=\frac{d\sigma }{dt}= \frac{4 \pi
\alpha^2 F_1^4 (t)}{t^2} - \frac{\alpha(\rho+ \Delta
\phi)\sigma_{tot} F_1^2 (t)}{|t|} e^{-B|t|/2}
 + \frac{\sigma_{tot}^2 (1+\rho^2 )}{166 \pi} e^{-B|t|},
\label{2.3}
\end{equation}
where the three terms are due to Coulomb scattering,
Coulomb-hadronic interference, and hadronic interactions. $L$ is
the integrated luminosity,  $\alpha$ the fine structure constant,
$\Delta\phi$  the relative Coulomb-hadronic phase, given as
\begin{equation}
\Delta\phi = -\ln(B(s)|t|/2)-0.577 \label{2.4}
\end{equation}
and $F_{1}(t)$ is the nucleon electromagnetic form factor, which
is usually parameterized as
    \begin{equation}
F_{1}(t) = \frac{1}{(1+|t|/0.71)^2}.
            \label{2.5}
\end{equation}
In the least squares fit procedure, the following two equations
are also used:

\begin{equation}
\sigma_{tot}^2 = \frac{1}{L}
\frac{16\pi}{(1+\rho^2)}\frac{dN_{el}}{dt}|_{t=0}; \label{2.6}
\end{equation}

  \begin{equation}
\sigma_{tot} = \frac{1}{L} (N_{el}+ N_{inel} ). \label{2.7}
\end{equation}

Eq. (\ref{2.6}) is a direct consequence of the Optical theorem.
$N_{el}$ is the total number of events obtained by integrating the
$dN_{el}/dt$ distribution within the $-t$ region where hadronic
interactions dominate, and extrapolated to $ -t = 0$ and to $t
\rightarrow \infty $ by using the form $\exp(-B |t|)$. $N_{inel}$
is the total number of inelastic events. Note that Eqs.
({\ref{2.6}) and ({\ref{2.7}) allow the luminosity $L$ to be
expressed in terms of $\sigma_{tot}$ and $\rho$.
 Then $dN_{el}/dt$ in Eq. (\ref{2.3}) can be expressed in terms of just three unknowns:
  $\sigma_{tot}$, $B$ and $\rho$. In the fit procedure, the same data on
$dN_{el}/dt$, together with the total number of inelastic events
$N_{inel}$ recorded during the same experimental data taking runs,
are used as inputs. A least-squared analysis for $\sigma_{tot}$,
$B$ and $\rho$ in Eq. (\ref{2.3}) is done by using all the
collected input data.

The evaluation of systematic errors due to the uncertainty in beam
emittance, vertex positions and spread, beam transport and
incoming beam angles is based on Monte Carlo and machine
simulations. These simulations use the geometry of the
experimental set-up and efficiency of the detectors as input.

\subsection{ Elastic scattering run scenarios}

For elastic scattering, three run scenarios are considered (Table
1):
\begin{enumerate}
\item   Nominal TOTEM optics for (low $-t$) elastic scattering,
$\beta^{*} \sim 1500 \ $m,

\item   An early medium- $\beta^{*}$ optics, with $\beta^{*}\cong
90 \ $m, and

\item   Optics for large $-t$ elastic scattering, $\beta^{*}= 2 -
0.55 \ $m.
\end{enumerate}
The event rate per bunch crossing is calculated as (for symbols see
Table 1)

\begin{equation}
 N_{ev} =\frac{\sigma_{el} \ L}{f \ N_{b}}\left(\frac{N_B}{N_b}\right),
            \label{1.29}
\end{equation}
where $\sigma_{el} = $ elastic cross section, $L = $ luminosity,
$f$ frequency, $N_{b}=$ no. of bunches; factor ($N_{B}/N_{b}
\approx 1$) accounts for the empty buckets.

\begin{table*}
\begin{tabular}{|c|c|c|c|} \hline
Scenario     & 1 & 2 &3   \\
     & low $|t|$ elastic,  &  low $|t|$ elastic,   &  large $|t|$ elastic,\\
  Physics:  & $\sigma_{tot}$(@$\sim 1 \% $), &  $\sigma_{tot}$(@$\sim 5 \% $),
  & hard diffraction    \\
    & MB, soft diffr. & MB, soft diffr. & \\ \hline
$ \beta^{*}$[m]  &  $\sim 1500$ & $90$ & $2 \div 0.5 $   \\
N of bunches  &  $43 \div 156 $ & 156 & $936 \div 2808 $   \\
Bunch spacing [ns] & $2025 \div 525 $ & $525 $ & $25 $   \\
N of part. per bunch  &  $(0.6 \div 1.15) \cdot 10^{11} $& $1.15 \cdot 10^{11} $
 & $1.15 \cdot 10^{11} $  \\
Half crossing angle [$\mu rad$]  & $0$ & $0 $ & $92 $   \\
Transv.norm.emitt. $\epsilon_n$[$\mu m rad$]  & $1$ & $3.75 $ & $3.75 $   \\
RMS beam size at IP [$\mu m$]  & $450$ & $213 $ & $32 $   \\
RMS beam diverg, at IP [$\mu rad$]  & $0.3$ & $2.3 $ & $16 $   \\
Peak Luminosity $[cm^{-2}s^{-1}]$
 &  $10^{28} \div 2 \cdot 10^{29} $& $3  \cdot 10^{30} $& $ 10^{33} $  \\ \hline
\end{tabular}
\caption{Three different LHC run scenarios foreseen for elastic
scattering measurements at the LHC. }
\end{table*}

\vspace{1cm}

\subsection{Total $pp$ cross section measurement strategy}

The total proton-proton cross section is measured - in a
luminosity independent way - by using the Optical theorem.
 By extrapolating the elastic rate down to the Optical point, $t = 0$,
 and by recording the elastic and  inelastic event rates, the total cross section
  is measured with an over-all accuracy better than $1 \% $.

For the total cross section measurement, the ``nominal'' TOTEM
beam optics ($\beta^{*} \sim  1500 \ $ m) with several short runs
is used. During the initial LHC running, the run scenarios with
$\beta^{*}\simeq 90 \ $m is planned to be used for a total cross
section measurement with an accuracy of about $5 \% $.

For measuring the total cross section, inelastic scattering needs
to be studied within large $E_T.$ The aim of the forward physics
initiatives at the LHC is to complement the base line ATLAS
(ALFA), CMS (TOTEM) and ALICE designs with forward detector
systems (see Refs.  \cite{2.5,2.6,2.7,Orava, 2.9} ). Besides
restricted detector acceptance, inadequate theoretical
understanding of the forward physics phenomena poses a serious
systematic uncertainty for the base line experiments in need of
precise luminosity measurement. As an example, the single
diffractive cross section for the low diffractive mass region, $
M^{*} < 5 \ $GeV, could amount to $25 \% $ of the over-all
$\sigma_{sd}$ and cause a major systematic uncertainty in the
total $pp$ cross section measurement.

The base line LHC experiments define a ``minimum bias'' event
category that must be  suppressed due to the limitations in
recording $pp$ interactions in excess of the rate ($1^{st}$ level
trigger band width) which can be recorded, as the cost of
triggering on ``interesting'' large $E_{T}$ events. The main task of
equipping the forward region of a main stream LHC experiment is to
complement the physics reach by including the events that are not
selected by the ``minimum bias'' event trigger. The TOTEM
experiment together with CMS forward detectors (CASTOR, ZDC, and
the proposed FSC and FP420 systems), represent the necessary
complement for selecting an {\it unbiased} (!) sample of ``minimum
bias'' events required for the analysis goals stated in the CMS
Physics TDR.

In general, the soft particles in the non-diffractive event
category (nd) will end up at central rapidities, while the
relatively few energetic ones are expected to end up at small
angles, to be recorded by the forward calorimetry and
spectrometers.

The longitudinal, $z$, coordinate of the vertex is defined by
first determining the distance of closest approach to the nominal
$z$ axis for each track, $z_{track}$, (at least two tracks are
required) and by then calculating the mean: $z_{vx} =
<z_{track}>$ . By using this simple method, the $z$ coordinate of
the primary interaction vertex can be determined with a resolution
of $\sigma_{vx} \sim 5 \ $cm.
 The beam related events are identified by requesting that their reconstructed
 $z_{vx}$ value is within  $\pm 50 \ $cm
 of the nominal IP. This selection is found to be more than $96 \% $
 efficient in choosing the beam related events.
 The remaining events have diffractively excited systems with relatively small masses,
 $M^{*} <   10 \ $GeV, in which all the charged tracks escape detection
 in TOTEM T1 and T2 spectrometers (Figure 4).

\begin{figure}[tbh!]
\begin{center}
\includegraphics[width=0.9\textwidth]{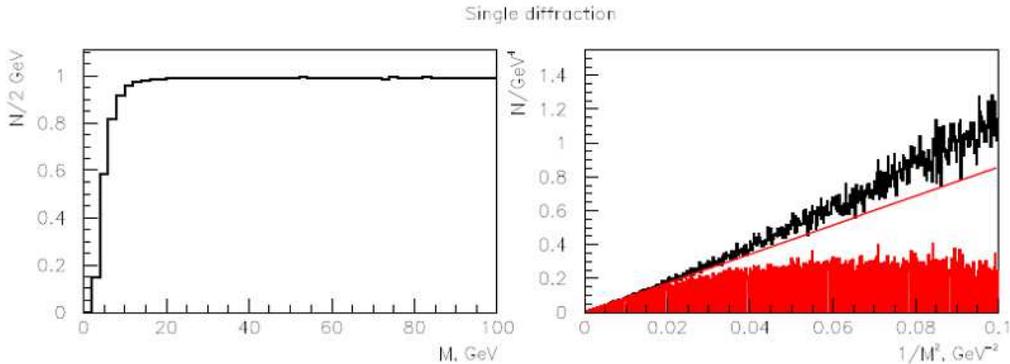}
\end{center}
\caption{Left panel: Ratio of detected single diffractive (sd) events as
a function of diffractive mass. Right panel: uncorrected Monte Carlo
simulated (unshaded) and acceptance corrected (shaded) sd events
with diffractive masses above 10 GeV \cite{2.1}.
 }
\label{fig:2.4}
\end{figure}

For obtaining the over-all inelastic rate, the missing
coverage\footnote{ The rapidity coverage could be extended further
by simple Forward Shower Counters (FSCs) placed at 60 to 140
meters from IP5. } above   $\eta >  7$ was estimated by
extrapolation.
 Fig. \ref{fig:2.4} (right panel) shows the simulated $1/M^{*2}$ distributions
 for the single diffractive ($sd$) events before and after acceptance correction.
 A linear fit is used to correct for the unseen part of spectrum.
 In the case of $sd$ events, a correction of
 $4 \% $, corresponding to about $0.6 \ $mb, was quoted above the detected fraction
 of this category of events.
 A similar analysis gives a correction factor of $0.1 \ $mb for
 double diffractive and $0.02 \ $mb for
 the central diffractive events \cite{2.1}.

 Unfortunately diffractive cross sections are poorly known and theoretical
 understanding of both small and large mass single diffraction is lacking.
 Low mass single diffraction ($M^{*} < 5 \ $ GeV) could represent $25 \% $
 of the total diffractive cross section (for double diffractive ($dd$) events
 the uncertainties are even more severe). Moreover, in the light of recent CDF measurements,
  soft central diffractive cross sections are likely to be seriously overestimated
  in current Monte Carlo models.

With the vertex constraint, a substantial part of the beam-gas
events is rejected. The simulation studies  \cite{2.1}  show that
by using the vertex constraint,
 the beam-gas interactions closer than
 $\pm 5 \ $m represent only  $\sim 3 \% $ of the selected sample of events.
 Since the beam-gas event rate is estimated to be $2 \ $ Hz/m at $1/20$
 of the nominal beam current,
 the trigger rate due to this source of background can be safely neglected.
 In addition, the Roman Pot based leading proton trigger $( \varepsilon \sim 90 \% )$
 can be used to further reduce the background rate.

 \subsection{$\sigma_{tot}$ measurement}

The LHC (TOTEM) measurements of the total $pp$ cross section is
luminosity independent
 and based on the Optical theorem:
\begin{eqnarray}
 \sigma_{tot}  =\frac{16 \ \pi}{(1+\rho^2)} \frac{(dN/dt)_{t=0}}{N_{el}+N_{inel}},
            \label{2.9}
\end{eqnarray}
where $N_{el}$ and $N_{inel}$ are the elastic and inelastic event
rates,
 ($dN_{el}/dt$)$|_{t=0}$
  is the elastic cross section extrapolated to the Optical point.

The relative error in $\sigma_{tot}$, neglecting uncertainty in
$\rho$ , is then:
\begin{eqnarray}
 \left(\frac{d \sigma_{tot}}{\sigma_{tot}}\right)^{2}=\left(\frac{d(dN_{el}/dt)_{t=0}}
 { (dN_{el}/dt)_{t=0} }\right)^{2}
 + \left(\frac{d(N_{el}+N_{inel})}{ (N_{el}+N_{inel}) }\right)^{2},
            \label{2.10}
\end{eqnarray}
The uncertainty in inelastic cross section  is estimated to be
less than 1 mb (see Ref. \cite{2.1}).
However, little experimental data on small mass diffraction
exists, and the uncertainty in inelastic diffraction alone could
amount to several millibarns. Together with the uncertainty of
about $ 0.5 \% $
of the extrapolated value of $(d\sigma_{el}/dt)|_{t=0} $, results
in an over-all error of
\begin{eqnarray}
  \frac{\Delta \sigma_{tot}}{\sigma_{tot}} >> 1 \% .
\end{eqnarray}
The uncertainty in the value of the  $\rho$-parameter could also
have an important contribution to the  $\sigma_{tot} $ measurement
 (see Chapter 3 and Ref. \cite{SelyuginPL94}), and it could be reduced by a direct measurement.

\subsection{Luminosity measurement and monitoring}
{\bf Luminosity measurement}

\vskip 0.5 cm

 The luminosity relates the cross section, $\sigma_{i}$, of a
given process $i$ to the corresponding event rate $ N_i$ by
\begin{eqnarray}
 L=\frac{N_i}{\sigma_i}.
            \label{2.11}
\end{eqnarray}
It is not trivial to find a process with a well defined - and
precisely calculable - cross section combined with a prominent
event signature. The cross section should be large enough for {\it
monitoring} the luminosity as a function of time, e.g. during a
fill or when investigating bunch-to-bunch variations.
 By using several complementary aprroaches, systematic uncertainties
 can be brought under control  \cite{2.3}.

A simultaneous measurement of the elastic and inelastic event
rates can be used to define the luminosity as
\begin{eqnarray}
 L  =\frac{1+\rho^2}{16 \ \pi} \frac{(N_{el}+N_{inel})^2}{(dN_{el}/dt)|_{t=0})},
            \label{2.12}
\end{eqnarray}

{\bf On-line luminosity monitoring}

\vskip 0.5 cm

 LHC is the first hadron collider where, due to the large c.m.s
energy and high luminosity,
 a significant number of inelastic interactions
 (on average 35 interactions at the design luminosity)
  are expected to take place per bunch crossing.
  The traditional technique of monitoring luminosity by
  requesting a {\it coincidence} of two counters at small angles on both sides
  of the IP is {\it not} sufficient at the LHC.
  At the design luminosity, the coincidence rate observed
  will not be proportional to the number of events,
   but rather to the number of {\it bunch crossings}.
   In the case of no segmentation of the luminosity monitors,
   the probability to get a coincidence will be close to $100 \% $.

Highly segmented forward detectors in  T2 and/or FSC
 could be used as {\it luminosity monitors}.
 The rate used to monitor the luminosity, could be defined by using
 the double-arm coincidence rate between
 a pair of left-right detector segments.
 The segmentation reduces the counting rate significantly below
  the bunch crossing frequency and, therefore,
  becomes {\it proportional} to the luminosity.
  A coincidence signal between a pair of left-right detector segments is unlikely in case
  of separate overlapping $pp$ collisions.
  The technique would help to suppress beam related backgrounds.
  In principle, beam related background may be recognized
   by a {\it time stamp} given by a forward detector.
   During the beam cross-over at the IP, no other bunches should pass through
    the luminosity monitor    location.
   Secondary particles from beam-gas and beam-wall interactions are traveling
   in-time with the bunch. In practice, the time stamping is challenging
   due to the high    bunch crossing frequency (40 MHz).
   By requiring simultaneous left-right signals contributions
   from the beam related backgrounds are practically eliminated.

The calibration of the luminosity monitors can be performed during
the dedicated high-$\beta^{*}$ runs at lower luminosities, where
the luminosity is precisely determined together with  tot.
Alternatively, once the total inelastic cross section is precisely
measured during the high- $\beta^{*}$ runs,
 together with the elastic and total cross sections,
 the inelastic events could be used to re-calibrate
 the monitor during the low-$\beta^{*}$ running.
 This becomes important when any significant changes to the detector
 lay-out are made,
 e.g. when the outer detectors are dismantled after the first year of running.

\section{Forward physics at the LHC} \label{Experiment}
Forward, or soft physics, roughly speaking, is the synonym of
diffraction - elastic and inelastic, the role of the latter
increasing with increasing scattering angle (or momentum
transfer).

Forward physics will play an important role in early runs of the
LHC for at least two reasons. One is that measurements of basic
quantities, such as the total cross section $\sigma_{tot}(s)$, the
ratio $\rho(s)$ of the real part to the imaginary one of the
forward scattering amplitude, the local slope $B(s,t),$ of the
differential cross section,  etc. are of fundamental importance
for the calibration/normalization of the beam and detectors and
this task goes beyond the problems of understanding the nature of
diffraction. Secondly, apart from classical studies of diffraction
(Pomeron), the diffractive medium (gluons) may also favour central
production of the Higgs boson. With increasing luminosity, the
experiments at LHC will gradually shift towards measurements of
rare events in the non-forward direction.

At the LHC, three collaborations, namely TOTEM/CMS, ALICE and
ATLAS, are preparing to measure elastic, inelastic and the total
$pp$ cross section \footnote {ALICE will use TOTEM's measurements
of the total cross section.} at the expected energy $\sqrt{s}=14$
TeV. The complementarity of the expected performances will ensure
optimal reliability of the results. The total cross section is
claimed to be measured within $1\%$ of precision. The precision of
the extrapolation to the optical point has been analyzed in
\cite{Helsinki1} \footnote{There are some reservations about the
achievable precision due to the uncertain cross section of
low-mass diffractive scattering that could amount to several
millibarns.}

One should note that the predictions based on model extrapolations
for $\sigma_{tot}^{pp}$ at $\sqrt{s}=14$ TeV have a wide range.
For example, Landshoff predicts \cite{Landshofftot_a,
Landshofftot_b} $\sigma_{tot}^{pp}$ (14 TeV) $= (125 \pm 25)$~mb.

TOTEM, CMS and FP420 collaborations are combining their
efforts to cover a phase space (see Fig. \ref{fig:lhcacceptance}),
where the geometric acceptance of detectors is shown) exceeding
that in any of the preceding collider experiment \cite{Orava,
Ferro, Deile}.


The TOTEM experiment, in particular, aims at measuring 1) the
total proton-proton cross-section with a relative precision of
about $1\%$; 2) the elastic proton-proton scattering between
$10^{-3}$ GeV$^2 <-t\approx(p\Theta)^2<10 $ GeV$^2$, $p$ and
$\Theta$ being respectively the proton momentun and scattering
angle in the c.m.s. Furthermore, by combining TOTEM with CMS and
its forward additions, the CASTOR and ZDC calorimeters, a full
forward physics program is planned.

The TOTEM experiment will measure $pp$ elastic scattering to about
$-t=0.15 \ $GeV$^2$ and will provide crucial new understanding of
the phenomena in the high $-t$ elastic scattering regime. To
discriminate between different models, it is important to measure
elastic scattering in the widest possible kinematic region. In
most of the previous measurements of the total cross section (at
ISR, SPS, Tevatron, RHIC) the value of the parameter $\rho$ was
imported either from another experiment or from model
calculations. As argued in Refs.~\cite{Deile, Sbarra}, known
variations of the value of $\rho$ have a negligible effect on the
resulting value of $\sigma_{tot}$. This is in disagreement with
Ref.~\cite{SelyuginPL05}, where it was argued that a small
variation of $\rho$ may affect the resulting $\sigma_{tot}$
significantly. If this is true, a simultaneous fit to all inputs
($\sigma_{tot}, \rho,$ and ${\cal L}$) from a single experiment is
needed. Publication of direct and unbiased data on ${sd-dN/{dt}}$
is highly welcome.

There are two possible kinds of luminosity measurements: one
yields an absolute value which serves as a point of reference, the
other one gives a relative value as a function of time. The latter
measurement will be performed by ATLAS using a special detector
called LUCID (Luminosity measurements Using Cherenkov Integrating
Detectors). The idea of the LUCID detector is explained e.g. in
Ref.~\cite{Royon}. It makes possible measurements in the very
forward region, directly related to the instantaneous luminosity.
This detector will enable ATLAS to obtain a linear relationship
between luminosity and the number of tracks counted in the
detector, directly related to the luminosity (see Table
\ref{table:Stenzel}).

Measurements in the Coulomb-nuclear interference (CNI) region,
$$-t\sim6.5\cdot10^{-4} \ {\rm GeV}^2,\ \ \Theta_{min}\sim 3.5 \; \mu rad,$$
(note that at SPS $\Theta_{min}\sim 120 \; \mu rad$) will be be
used to extract the value of the parameter $\rho$ from Eq. (8).

The extracted value of $\rho$ may be affected by at least two
phenomena. One is connected with the well known corrections
\cite{Kundratetal} to the Coulomb-hadron phase \cite{WY} and the
other one with the non-exponential behavior of the diffraction
cone, known as the ``fine structure of the Pomeron trajectory"
\cite{Shelk_a,Shelk_b,Shelk_c,Gribov_a,Gribov_b,Gribov_c,Gribov_d}.
Below, in Sec. 4, we shall come back to this point. The program of
studying this ``fine structure of the Pomeron trajectory" is among
the priorities of the ATLAS collaboration \cite{Sbarra,Royon}.

The ALICE experiment is designed as a general purpose experiment
with a central barrel covering the pseudorapidity range
$-0.9<\eta<0.9$ and a muon spectrometer covering the range
$-4.0<\eta<-2.5$ at luminosities  ${\cal L}=5\cdot
10^{30}cm^{-2}s^{-1}$ and ${\cal L}=10^{27}cm^{-2}s^{-1}$ in $pp$
and $PbPb$ collisions, respectively, as well as an asymmetric
system $pPb$ at a luminosity ${\cal L}=10^{29}cm^{-2}s^{-1}$
(see, also, Ref.~\cite{Schicker_a,Schicker_b,Schicker_c}).

Moreover, the experimental program of ALICE to large extent will
be oriented to inelastic reactions, e.g. by studying their
dependence on the width of the rapidity gap.

It should be noted that all these values are approximations since
some of them were taken
 at $t=0$ as $\rho(0)$, $dN_{el}/dt\mid_{t=0}$, or integrated over
all the region of angles as $N _{el}$, $N_{inel}$, i.e. all these
values were obtained under some theoretical assumptions.

\begin{table*}
      %
\begin{tabular}{llll}
\hline\noalign{\smallskip}
        &  input &   fit & stat.error \\
\hline\noalign{\smallskip}
  $ {\cal {L}}$  &   8.10 \ 10$^{26}$  &  8.151  \ 10$^{26}$   &   1.77 \%  \\
  $\sigma_{tot}$  & 101.5  \  mb & 101.14  \  mb & 0.9\%  \\
   $B $ & 18 GeV$^{-2}$  &    17.93 \  GeV$^{-2}$  & 0.3\%   \\
  $\rho$   &  0.15  &  0.143  & 4.3 \% \\
\noalign{\smallskip}\hline
\end{tabular}
\caption{Result of a fit (see Ref.~\cite{Stenzel}) to simulated
$dN/dt$ data corresponding to $\sim$ 1 week ($10\ M$ events) of
running.
${\cal L} =10^{27} cm^{-2}s^{-1}$.\label{table:Stenzel}}
\end{table*}

\begin{table*}
\begin{tabular}{lllllll}
\hline\noalign{\smallskip}
       $\sqrt{s}$ &   $\sigma_{tot}$ &   $\delta \sigma_{tot}$ &
Authors& publication \\
\hline\noalign{\smallskip}
   540 & 66.800 &   5.90  &  ARNISON 83 &  PL 128B, 336  \\
   541  & 66.000 &  7.00  & BATTISTON 82 &     PL 117B, 126 \\
   546  & 63.000 &   2.10   & AUGIER 94 &        PL 344B, 451   \\
   547  &  61.260 &   0.93   & ABE 93S       &   PR D50, 5550  \\
   900 &  61.90   & 1.50 &  BOZZO 84B     &   PL 147B, 392 \\
  1800  & 65.30  &   0.70      &  ALNER 86  &       ZP C32, 153 \\
  1800  & 71.42 &   1.55    & AVILA 02    &     PL 537B, 41      \\
  1800  & 80.03 &  2.24   & ABE 93S   &      PR D50, 5550 \\
  1800  &  72.80 &   3.10  & AMOS 91B  &      PRL 68, 2433 \\
\noalign{\smallskip}\hline
\end{tabular}
\caption{Range of the measured values of $\sigma_{tot}$ above the
ISR energies
\label{table:sigmatot}}       %
\end{table*}

Tables \ref{table:sigmatot} and \ref{table:ratiorho}
\cite{SelyuginEDS07} show the range of the measured and the
expected values of $\sigma_{tot}$ above the ISR energies, with the
divergence of the existing data on $\sigma_{tot}$ at the same
energy quoted in Table \ref{table:Stenzel}. One can see that the
extraction of $\sigma_{tot}$ from the differential cross section
$dN/dt$ is a complicated problem and that the use of different
models can chance the predicted value of the total cross sections
at the LHC energies from $80$ mb to $230$ mb (see Table 3).
Therefore, one cannot use the theoretical predictions of, say
$\sigma_{tot}$ or $\rho$, to extract other observables from the
experimental data on $dN/dt$. With the exception of the UA4 and
UA4/2 Collaborations, the numerical data on $dN/dt$ of other
collaborations were not published, excluding any cross-check or
improvement of these results. We hope that the future LHC data on
$dN/dt$ will be published.

 In Ref.~\cite{Royon}
the significant correlations between the values of $\rho$ and
$\sigma_{tot}$ are also visible.
   The results of the fits of the simulated LHC experimental data in
the framework of the non-exponential model of the hadron
 scattering amplitude presented at ``EDS-07"  \cite{SelyuginEDS07}
  at $\sqrt{s}$ equal to $2$~TeV and $14$~TeV  show large errors in the
  determination of $\sigma_{tot}$ and  $\rho(0)$
(see Tables \ref{table:sigmatot} and \ref{table:ratiorho}).

\begin{table*}
\begin{center}
\begin{tabular}{llllll}
\noalign{\smallskip}\hline\noalign{\smallskip}
Collaboration & & $\sigma_{tot}$ (mb) & $\sigma_{el}/\sigma_{tot}$ & $\rho(t=0)$ & $B(t=0)$ \\
\hline\noalign{\smallskip} \hline\noalign{\smallskip}
  KMR & \cite{KMR} & 88.0 (86.3) &0.22 (0.209)   & -  & - \\
  C. Bourrely et al & \cite{BSW} & 103 &  0.28  & 0.12  & 19 \\
  E. Gotsman et al & \cite{GLM} & 110.5  & 0.229    &  - & 20.5  \\
  COMPETE Coll. & \cite{compete} &  111 &  -  &  0.11 & -  \\
  B. Nicolescu et al & \cite{Paris} & 123.3  &  -  & 0.103  &  - \\
  S. Goloskokov et al & \cite{SelyuginPN87_a,SelyuginPN87_b} & 128  &  0.33 &  0.19  &  21   \\
  J.R. Cudell et al & \cite{SelyuginBDL06}  &  150 & 0.29 & 0.24  & 21.4 \\
  Petrov et al & \cite{Protvino} &  230   & 0.67 & -  &  - \\
V. Petrov, A. Prokudin & \cite{Prokudin03} & 107 & 0.28 &  0.138 & - \\
M. Islam et al & \cite{Islam} & 110 & - & 0.12 & -\\
\noalign{\smallskip}\hline
\end{tabular}
\caption{ Predictions of the parameters of elastic scattering
amplitude at $ \ (\sqrt{s} =14 \ $TeV, $t=0$)
\label{table:ratiorho}}
\end{center}
\end{table*}

The (weak) dependence  of $\sigma_{tot}$ on $\rho,$ if $\rho$ is
very small, and it comes from the coefficient $1/(1+\rho^2)$ in
front of $d\sigma/dt$ (see Eq.~(\ref{2.9})), while the strong
dependence of the normalization of $dN/dt$, the values of
$\sigma_{tot}$ and $\rho(0)$ comes from the extraction of the
Coulombic and Coulomb-hadron interference terms from  $dN/dt$ data
to obtain $(d \sigma_{el}/dt)_{t=0}$. The Coulomb-hadron
interference term is proportional to $[ \rho(s,t) - \alpha
\Phi_{cn}(s,t)]$. At very small $t,$ the Coulomb-hadron phase is
also important. Corrections to the Bethe formula~\cite{Bethe} were calculated
in Refs.~\cite{Cahn82,selmp1,selmp2,selprd}. A detailed analysis
of the role of this term in the behaviour of the differential
cross sections was carried out in Ref. \cite{Prokudin03}. To make
the analysis complete, the Odderon contribution
\cite{Shelk_a,Shelk_b,Shelk_c} as well as the nearby
threshold singularity at $t=4m_{\pi}^2$ \cite{Gribov_a}, should be
also taken into account.

  To minimize the errors, in some experiments the value of $\sigma_{tot}$
or $\rho$
  were fixed from other measurements. This was done, for example, by the
UA4/2 Callaboration,
  which extracted $\rho(0)$ by using $\sigma_{tot}$ from  the result
  of the UA4 Collaboration ($\sigma_{tot}=61.9 $ mb).
  However, the value of $\sigma_{tot}$ obtained by the Collaboration  UA4/2
turned out to be $\sigma_{tot}=63.0 $.
 With such a value of $\sigma_{tot} $ the resulting $\rho(0)$ becomes larger.
This situation was
 analysed in Refs.~\cite{SelyuginYF92, SelyuginPL94}.
 with the results shown in Table~\ref{table:rho2}.

 Diffraction dissociation, in particular the low mass one, is among priorities of
 the first experiments at the LHC. It should be remembered that diffraction
 dissociation - both single and double - and elastic scattering have
 much in common. Similarities are expected in the shape of the diffraction cone, with its fine
 structure (the ``break'' and oscillations), open for observation, see Ref. \cite{Goulian},
 and the dip-bump structure (until now not seen in diffraction
 dissociation).
Hence studies of elastic scattering and diffraction
 dissociation are complementary. A
recent overview of the ALICE detector and trigger strategy for
diffractive and electromagnetic processes at the LHC can be found
in Ref. \cite{Schicker_a,Schicker_b,Schicker_c}.

\begin{table*}
\label{comparison-of-rho}       
\begin{tabular}{lllll}
\hline\noalign{\smallskip}
\multicolumn{5}{c} {${\rho}  \ (\sqrt{s} =540 \ $GeV, $0.000875 \leq |t| \leq 0.12 \ $GeV$^2$)}   \\
\noalign{\smallskip}\hline\noalign{\smallskip}
n & experiment      & exp. analysis      & mod. anal. I
\cite{SelyuginYF92} &mod. anal. II \cite{SelyuginPL94}  \\

1 & UA4      & $0.24 \pm 0.02$      & $0.19 \pm 0.03$ & - \\

2 & UA4/2    &  $0.135 \pm 0.015$     & - & $0.17 \pm 0.02$    \\

\noalign{\smallskip}\hline
\end{tabular}
\caption{Comparison of experimental values of the ratio $\rho$ of
the real to imaginary part of the scattering amplitude, obtained
in the UA4 and UA4/2 experiments, with theoretical values.
\label{table:rho2}}
\end{table*}

\section{The forward cone\label{s3}}
\subsection{The ``break'' at very small $|t|$}
An essential part of the future TOTEM+CMS and ATLAS experiments
is connected with the measurement of the  elastic differential
cross section at small momentum transfer, with the purpose of extracting
from these data the values of the total cross section. An
important point of this procedure is the simultaneous measurement
of four quantities: the luminosity  $L$ (or the normalization
coefficient), the total cross section $\sigma_{tot}$, the slope
$B(s,t)$, defined as
\begin{equation}
B(s,t)= {\frac {d}{dt}} \log {\frac {d\sigma(s,t)}{dt}}
\end{equation}
and the ratio $\rho$. The last two quantities depend on their, a
priori unknown, $t$-dependence. Consequences of this complexity
are the contradictions of the obtained values of $\rho(s)$ in
different experiments in the energy range of $\sqrt{s} = 5 \div 20
\ $GeV \cite{SelyuginYF92}, in particular the large difference in
the values of $\rho$ as measured in the UA4 and UA4/2 experiments
at $\sqrt{s} = 540 \ $GeV as well as the discrepancy between the
values of $\sigma_{tot}$ at $\sqrt{s} = 1.8 \ $TeV. This
discrepancy results from the correlations between simultaneous
measurements of several unknown values in a single experiment
instead of fixing some values from the phenomenological analysis
or from other experiments.

This situation was demonstrated for example in
Ref.~\cite{SelyuginYF92}. Since we do not know the true
$t$-dependence of $\rho$ and of $B$, one cannot use the same
constant values for different small intervals of $t$. Therefore,
simultaneous measurements of all unknown observables in a single
experiment are needed, rather than their substitution by values
taken either from phenomenological models or alternative
experiments, as illustrated e.g. in Refs. \cite{SelyuginYF92,
SelyuginPL94}, where compatible values of $\rho$, different from
those experimental were obtained.

Given the uncertainty in the $s$ and $t$ dependence of the
observables, the standard procedure is to suppose their smooth and
monotonic behavior from low to super-high energies. Along these
lines, it is usually proposed to measure $\sigma_{tot}$ taking the
value of $\rho$ from the results of theoretical analysis (for
example, that of the COMPETE Collaboration \cite{compete}),
although, as argued in Refs.
\cite{Helsinki_a,Helsinki_b,Helsinki_c,Helsinki_d}, any consistent
analysis should include a simultaneous fit to expression of
Eq.~(\ref{2.3}).


In the ISR energy region, the $pp$ diffraction cone changes its
slope $B(s,t)$ near $t=-0.1$ GeV$^2$ by about 2 units of
GeV$^{-2}$. In Ref.~\cite{Gribov_a} this phenomenon was
interpreted as the manifestation of $t-$channel unitarity (a
two-pion loop, see Fig.~\ref{two-pion}) and, in term of the
Pomeron exchange, it was modeled by the inclusion of a relevant
threshold singularity in the trajectory\footnote{This threshold
singularity can be included also in the logarithmic trajectory as
in Eq. (\ref{combined}).}:

\begin{eqnarray}
 \alpha_P(t)=\alpha_0+\alpha_1 t-\alpha_2\sqrt{4m_{\pi}^2-t},
\end{eqnarray}
where $\alpha_2/\alpha_1\approx 0.1.$ It will be interesting to
see whether this effect will persist at the LHC and thus confirm
the idea of Ref.~\cite{Gribov_a}.

As shown in Ref. \cite{Goulian}, the ``break", or
the fine structure of the Pomeron, in principle, can be seen
directly, unbiased by electromagnetic interactions, in
proton-neutron scattering.

\subsection{Oscillations}

An important phenomenon that should be scrutinized at the LHC is
the possible appearance of small-period oscillations over the
smooth exponential cone. Possible deviations from a simple
exponential behaviour of the hadron-hadron scattering amplitude
were discussed in the literature long ago. For example, it was
shown in Ref. \cite{Zarev} that peripheral contributions from
inelastic diffraction result in large- and small-period
oscillations in the momentum transfer. Among the attempts to
verify these oscillations the first one was done in an experiment
at Serpukhov, where oscillations in $pp$ scattering were
detected~\cite{Antipov} and then discussed in Ref.~\cite{Zotov}.
An alternative view, interpreting these oscillations as an
artifact, connected with the $t-$ dependence of the slope, was put
forward in Ref.~\cite{Selyugin85}. In Ref.~\cite{SelyuginProt82}
the statistical nature of the possible oscillations in the ISR
data were analyzed in the framework of the Dubna Dynamical (DD)
model (for the DD model see Subsection~\ref{sec:ddmodel}). The
results are shown in Fig.~\ref{fig:oscilation}.

The effect was first observed at the ISR and subsequently was
discussed in Ref.~\cite{Lengyel_a,Lengyel_b}. Although it has not
yet been confirmed unambiguously, it continues to attract
attention (see Ref.~\cite{Lengyel_a,Lengyel_b} and references
therein). The analysis of the slope \cite{Lengyel_a,Lengyel_b}
shows possible oscillations in the UA4/2 data. The super-fine
structure (oscillations) of the cone may be related to residual,
long-range interaction between nucleons \cite{Kuraj} or the action
of a potential of rigid hadronic strings \cite{Selyugin95}.

\begin{figure}[tbh!]
\begin{center}
\includegraphics[width=0.9\textwidth]{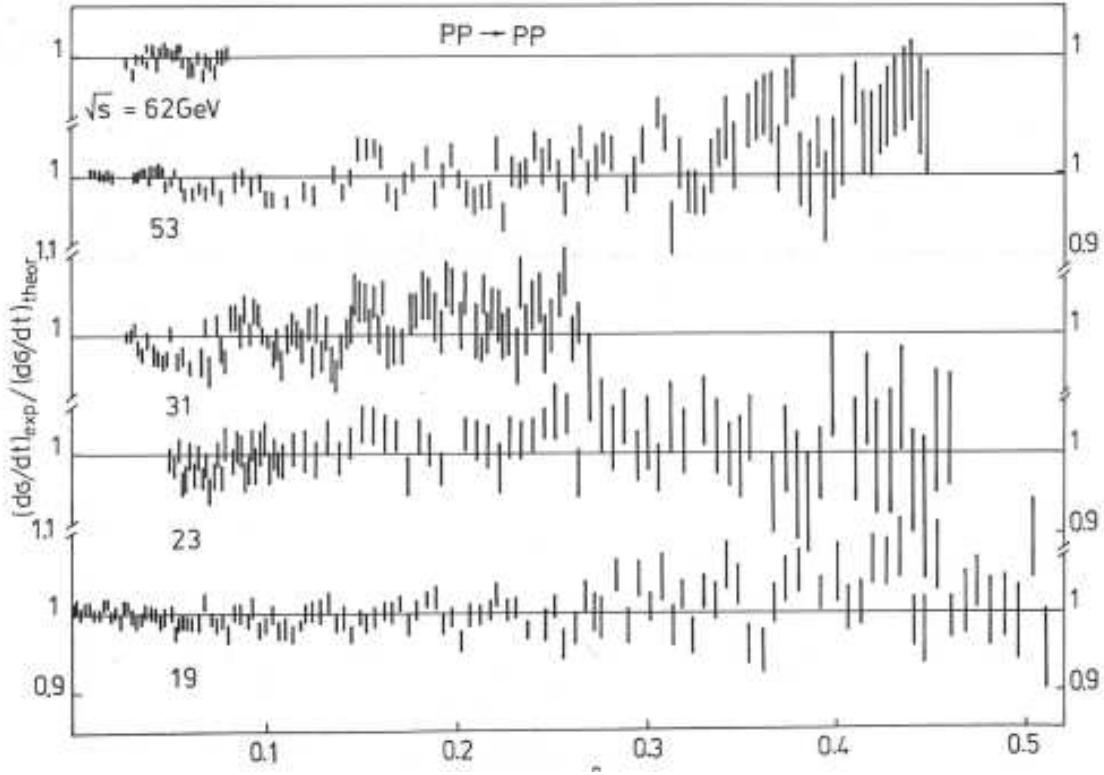}
\end{center}
\caption{Ratio of the experimental differential cross section to
theoretic one calculated from the DD model \cite{SelyuginProt82}
in the ISR energy region.}
\vspace{0.5cm}
\label{fig:oscilation}
\end{figure}

A new method of the analysis of the experimental data of UA4/2 experiment
    was proposed in Ref. \cite{SelyuginPL97}.
  The method is based on the
 comparison of two statistically independent sets,
Ref. \cite{hud}.
   If we have two statistically independent sets
  $x^{'}_{n_1}$  and $x^{"}_{n_2}$
  of values of the quantity  $X$  distributed around
  a certain value, say $A,$ with the standard error  equal to $1$,
  we can try
  to find the difference between
  $x^{'}_{n_1}$  and $x^{"}_{n_2}$.
  For that we can compare
  the arithmetic mean of these choices:
$$ \Delta X = (x^{'}_1 + x^{'}_2 + ... x^{'}_{n1})/n_1  -
        (x^{"}_1 + x^{"}_2 + ... x^{"}_{n2})/n_2  =
     \overline{x^{'}_{n_1}} - \overline{x^{"}_{n_2}}.   $$
  The standard deviation for this case will be
$$   \delta_{\overline{x}} = [1/n_1 +1/n_2]^{1/2}. $$

If we have the purely statistical noise the value  $\Delta X /
\delta_{\overline{x}}$
 tends to zero. However if there is some additional signal,
 this value should differ from zero.
 When it
is larger than $3$, one can say
 that the difference between these two choices has a
  $99\%$ probability of confirming the presence of the oscillations
 (for more details see  Ref. \cite{SelyuginPL97}).

The deviation $\Delta R_i$ of each of the experimental data
for the cross section from the corresponding theoretical values
 is measured in units of the experimental error $\delta_{i}^{exp} $:
  \begin{eqnarray}
\Delta R_ {i} = [(d\sigma/dt_{i})^{exp}
  -  (d\sigma/dt_{i})^{th}] / \delta_{i}^{exp}.
  \end{eqnarray}

 By summing these $\Delta R_{i}$ over all 99 experimental
points of the UA4/2 experiment, the
 result should tend to zero as the statistical deviations are equally
 probable in both sides of the theoretical curve. However,
 if the theoretical
 curve does not precisely describe the experimental data,
 for example, the scattering amplitude deviates from the
 exponential behavior in the momenta transfer, the sum over $\Delta R_{i} $
 can differ slightly from zero, going beyond the value
 of a statistical error.
 To take into account this effect, we divide
 the whole interval of the momentum transfer into $k$  equal pieces
 of size $\Delta$ such that $k \Delta \geq (q_{99} -q_{1}) $,
where $q_i = \sqrt{|t_i|}$, and
  sum  $\Delta R_{i} $
 separately over the even and odd pieces.
 Thus, we get two sums $L^{up} $ and $L^{dn} $ for the
  $n_1$ even and $n_2$ odd interval, respectively. For
  $n_1 + n_2 = k$ and $|n_1 - n_2| = 0$ or $1$:
    \begin{eqnarray}
   L^{up} = \sum_ {j=1}^{n_1} (\sum^{99}_{i}
            \Delta R_{i})|_{\Delta q (2j-1) < q_i \leq \Delta q(2j) }; \ \ \
   L^{dn} = \sum_{j=1}^{n_2} (\sum^{99}_{i=1} \Delta R_{i})
|_{\Delta q (2j) < q_i \leq  \Delta q (2j + 1) }.
  \end{eqnarray}

 In Ref.~\cite{SelyuginPL97}, where this method  was applied
 to the data of the UA4/c Collaboration,
 $\Delta q =0.9085 \cdot 10^ {-2} $ GeV was used.

 Let us calculate the quantities $L^{up} $ and $L^{dn} $;
 the results are shown in
 Fig.~\ref{fig:sums} by the full and dash-dotted lines. It can be seen that
in the range $0 < |t| \leq 0.1$ GeV$^2$ these quantities change drastically and
in the range $|t| > 0.1$ GeV$^2$ instead they vary slightly.
It means that the amplitude of a possible
 periodic structure is decreasing with growing $t$.

\begin{figure}[tbh!]
\includegraphics[width=0.5\textwidth]{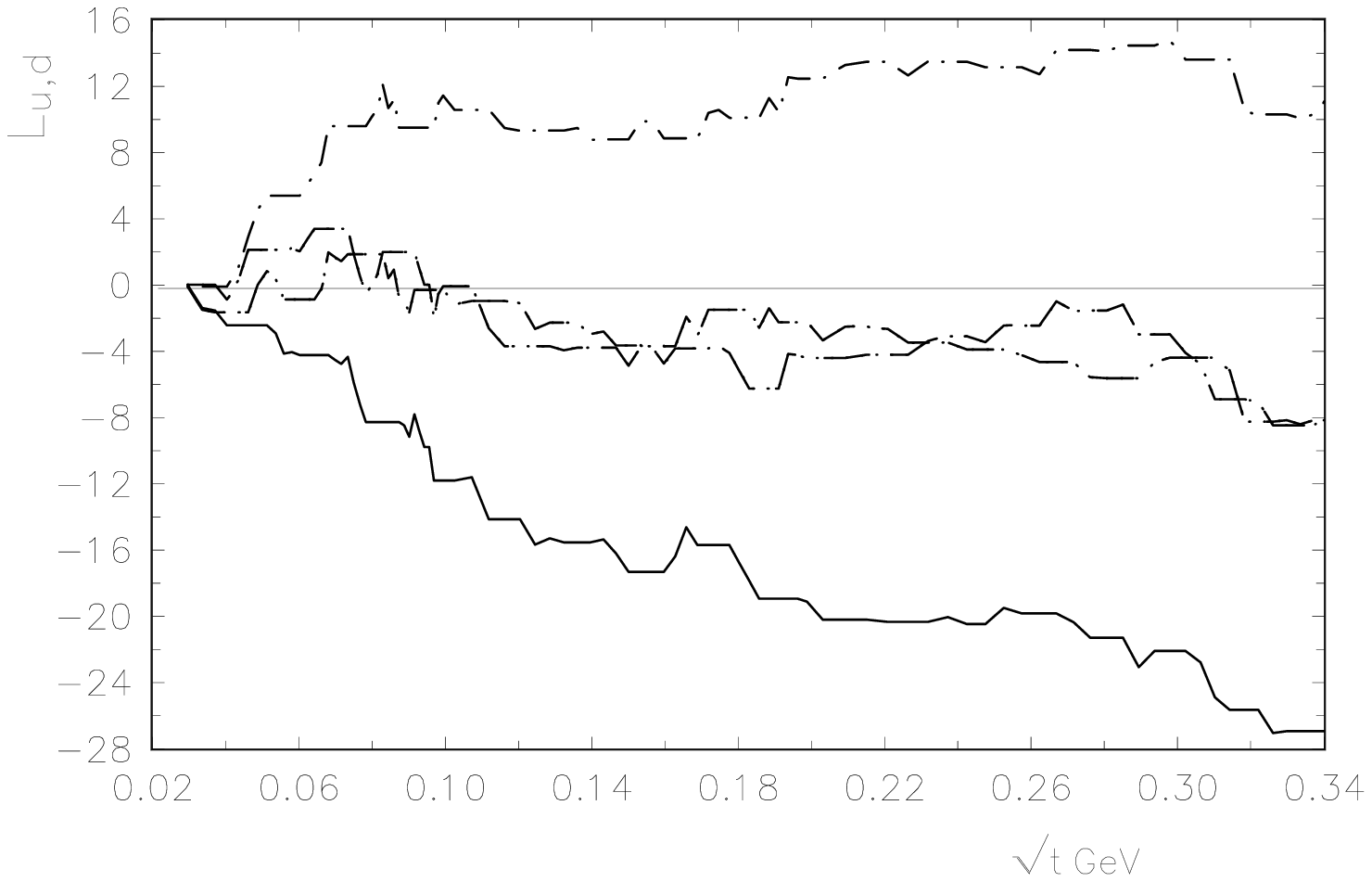}
\includegraphics[width=0.5\textwidth]{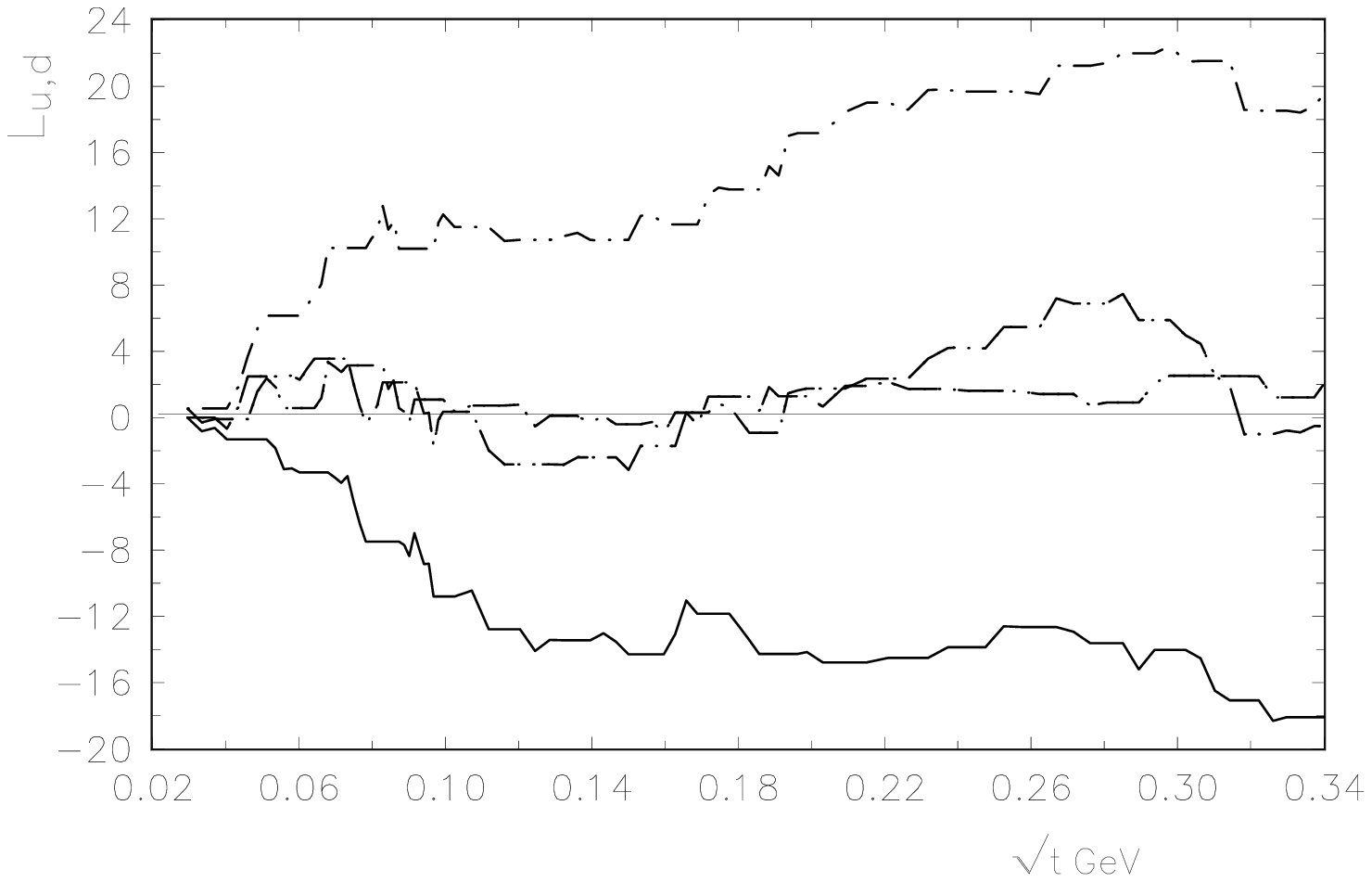}
 \caption{Left
panel: Sums $L^{up}$ and $L^{dn}$ calculated \cite{SelyuginPL94}
with $ \sigma_{tot}= 62.2 \ mb; \  B=15.5 \ GeV^{-2}; \
\rho=0.135 $
 for $q_{0}$ and  $\delta q $ (full and dashed lines);
 and for   $q_{0} + \delta q /2 $ ( dots and dots-dashed lines).
  Right panel: Same as on the left panel but with
$ \sigma_{tot}= 63.54 \ mb; \  B=15.485 \ GeV^{-2}; \  \rho=0.158
$.}
\label{fig:sums}
\vspace{0.5cm}
\end{figure}

Note that this new method can be used to check the true
determination of the parameters of the elastic scattering at small
$t$. The two curves obtained have to be symmetric with respect to
the line calculated by using the basic parameters ($\sigma_{tot},
\rho, B$).

\begin{figure}[tbh!]
\includegraphics[width=0.5\textwidth]{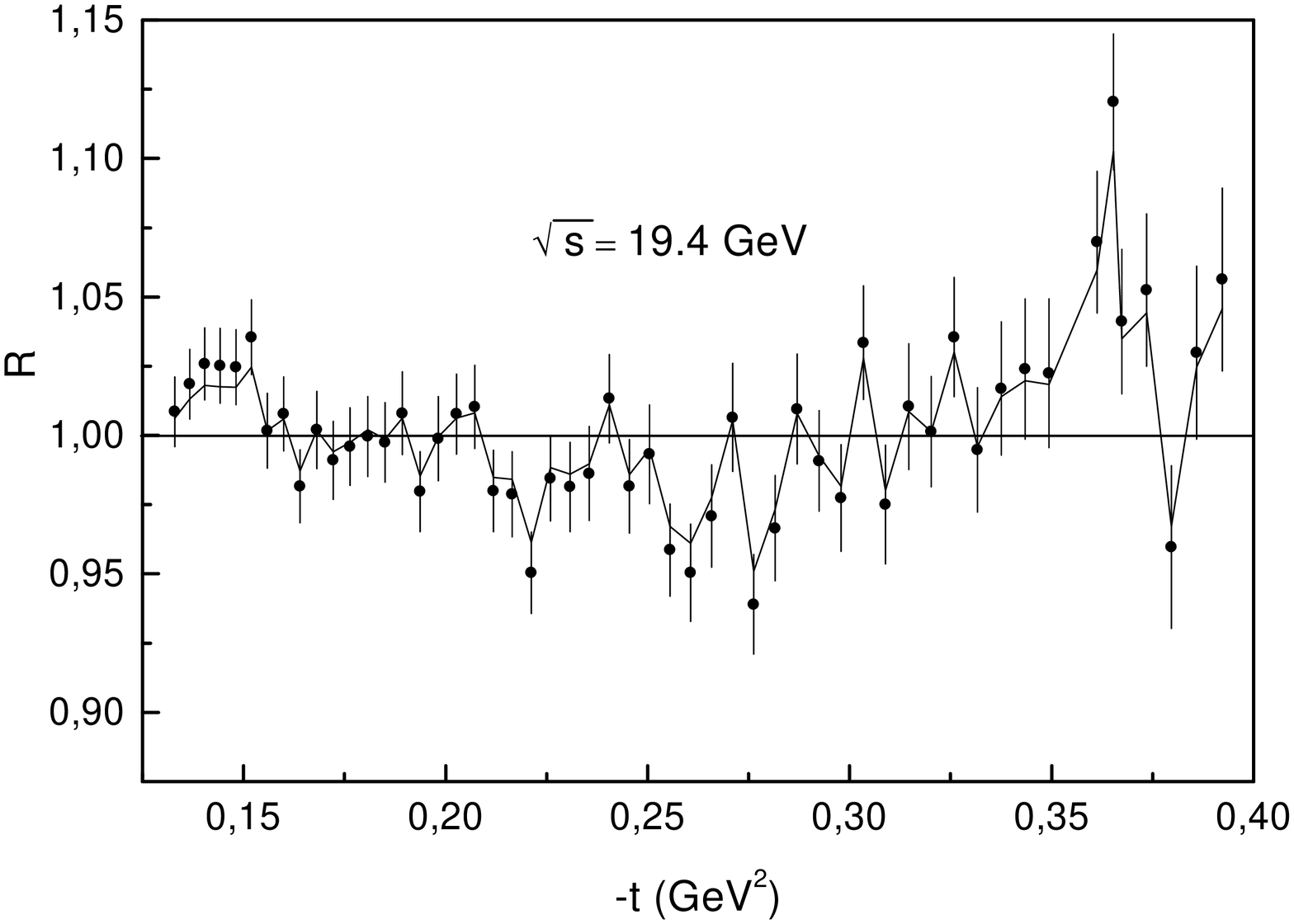}
\includegraphics[width=0.5\textwidth]{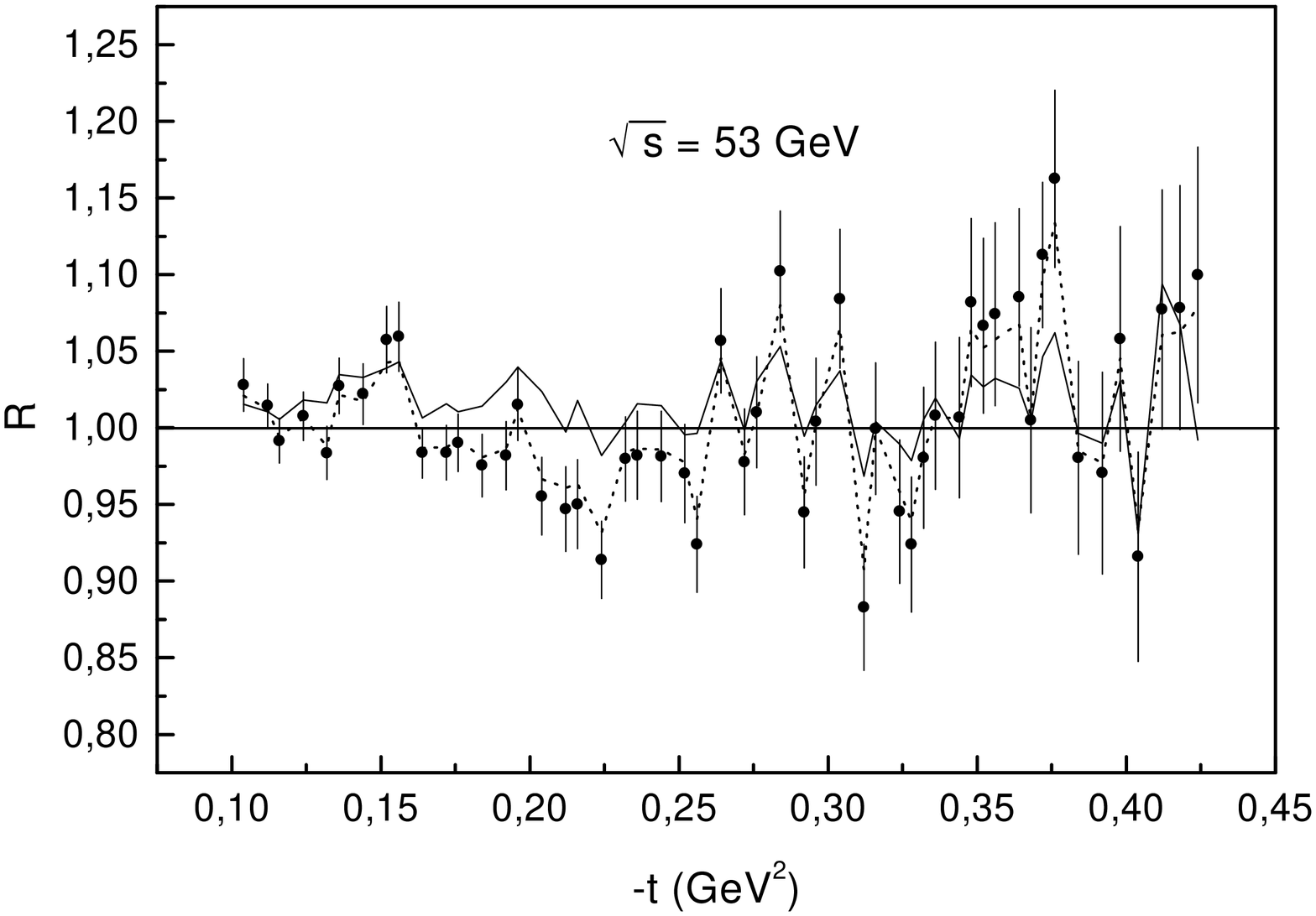}
\caption{Left panel: $t-$ oscillations calculated in
Ref.~\cite{Lengyel_a} as the ratio of the experimental to the
theoretical values at $\sqrt{ s}=19.4$ GeV. Data are from
Ref.~\cite{Schiz}. Right panel: Same as on the left panel but for
the data from Ref.~\cite{Barbiellini}.} \label{fig:oscill1.eps}
\vspace{0.5cm}
\end{figure}

A way to identify possible oscillations at the LHC is to use the
method of overlapping bins, suggested by J. Kontros and Lengyel in
Ref.~\cite{Lengyel_a}. The procedure consists in scanning the cone
by overlapping bins in $t$, each containing a certain number $N$
of data points shifted by a small number of points $n$. Within
each bin an exponential fit is applied~\cite{Lengyel_a,Lengyel_b}
(see Fig.~\ref{fig:oscill1.eps}).

The length of the bin should not be too large to oversimplify the
parameterization and not too small in order to contain a
reasonable number of points for each process. For example, a shift
from bin to bin $\delta t=0.01$ GeV$^2$ was found in Ref.
\cite{Lengyel_a} reasonable.

Looking for oscillations in the cone at LHC by using the method of
overlapping bins of Ref. \cite{Lengyel_a,Lengyel_b} is a promising
program for future experiments.  As already mentioned, the
oscillations can be related to residual long-range forces between
nucleons \cite{Kuraj}.

\section{The Dip-Bump Region, $t\sim -1$ GeV$^2$} \label{s4}

Before going into details, we would like to notice a
model-independent regularity found in Ref.~\cite{SelyuginNP87}. As
shown in that paper, a correlation between the value of $\rho$ and
the depths of the minima of the diffraction differential cross
section in proton-proton and proton-antiproton scattering exists.
The ratio $\rho$ for the latter changes sign in the energy region
$\sqrt{s} = 9.8 \ $GeV. At this energy the differential cross
section of $p\bar{p}$-scattering in the dip-bump structure has its
sharpest minimum, while  $\rho$ in $pp$-scattering changes its
sign around $\sqrt{s} \sim 30 \ $GeV. At this energy the
differential cross section of $pp$-scattering has its sharpest
minimum.

  At the highest ISR energies all models and experiments show that
$\rho(0)_{\bar{p}p} > \rho(0)_{pp}$. The experiments also show
that the dip at these energies in $\bar{p}p$ is higher then in
$pp$-scattering.

The observed dip-bump structure in the high-energy differential
cross section was not predicted by any model or theory. It can be
related e.g. to the multiple scattering of quarks and gluons
(Glauber theory), leaving however much room for speculations on
the basic inputs in this approach.  The optical model (see Ref.
\cite{Yang} and earlier references therein) predicted a sequence
of minima and maxima, however the ISR and SPS data show only one
structure, both in $pp$ and in $p\bar p$ scattering. Some models
\cite{Predazzi} predict that more structure will appear at higher
energies, for example at the LHC. The (non)appearance of additional
minima and maxima at the LHC will confirm or rule out part of the
existing models, although most of them will survive by refitting
its parameters a posteriori.

Two models \cite{Nicolescu, Soffer_a,Soffer_b}, fitted to the data
in a wide range of $s$ and $t,$ can illustrate the state of art in
this field (see, for instance, Fig.~\ref{fig:Soffer1.eps}). While
the authors of Ref.~\cite{Nicolescu} claim that fits require the
presence of an Odderon contribution, those of Ref.~\cite{Soffer_a,Soffer_b}
do not need it. By this we only want to stress that the
flexibility of the existing models allow for good ``postdictions",
but their predictions are not unique. In this situation useful
empirical parameterization \cite{Menon}, unbiased by any
theoretical prejudice can be useful.

\begin{figure}[tbh!]
\includegraphics[width=0.46\textwidth]{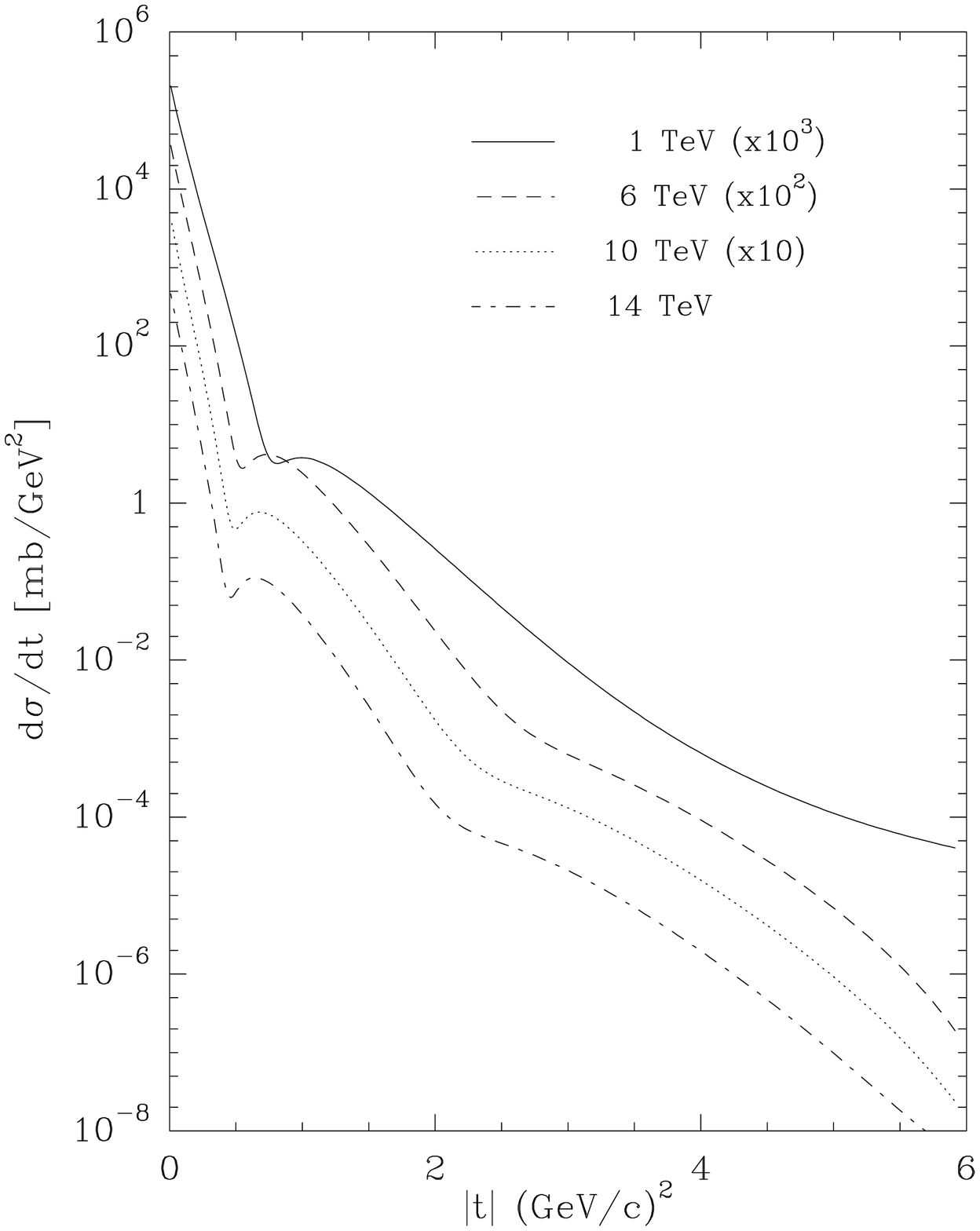}
\includegraphics[width=0.5\textwidth]{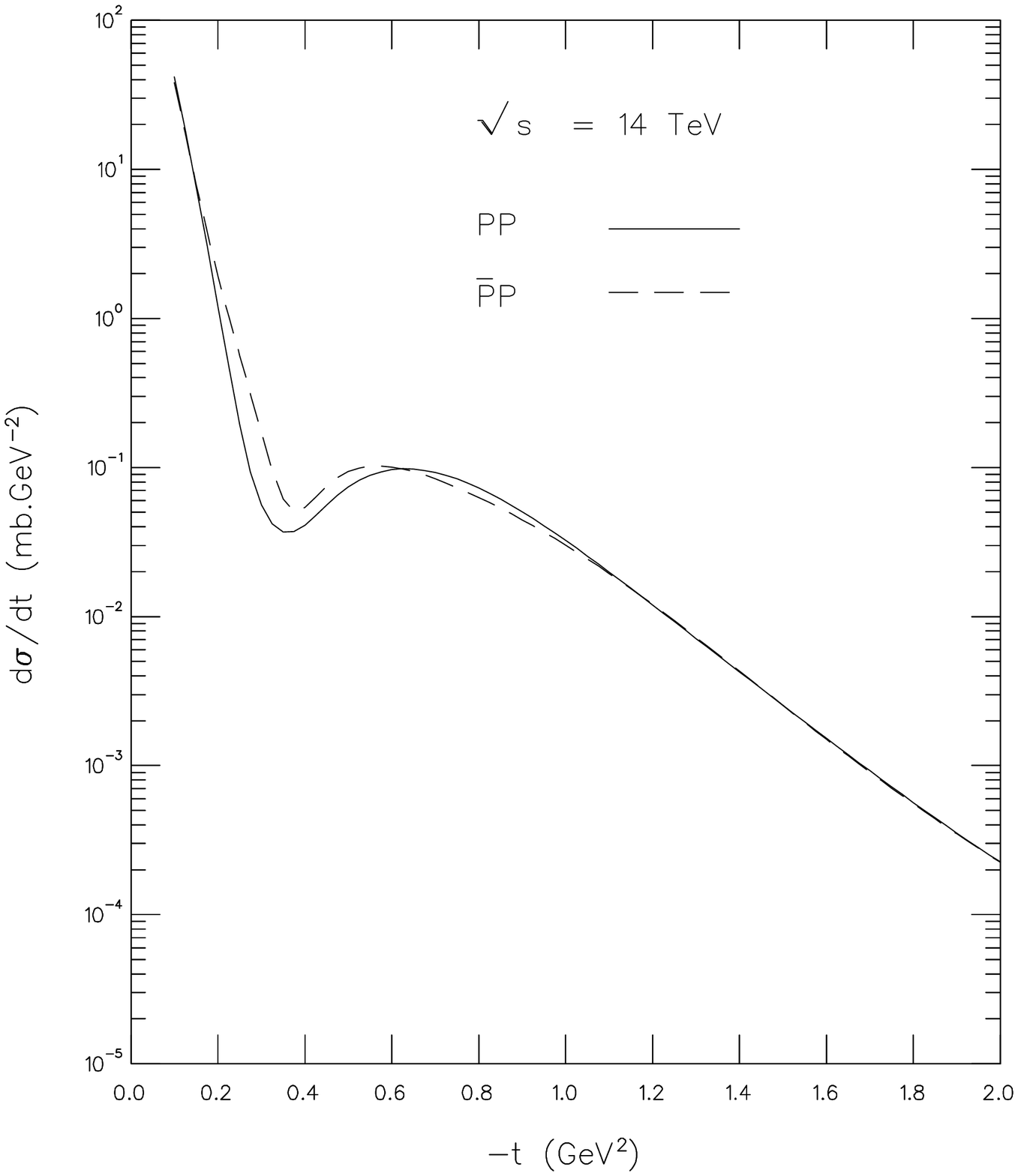}
 \caption{Left
panel: Elastic $pp$ differential cross section predicted in
Ref.~\cite{Soffer_a,Soffer_b}. Right panel: Elastic $pp$ and
$p\bar p$ differential cross section predicted in
Ref.~\cite{Nicolescu}.} \label{fig:Soffer1.eps} \vspace{0.5cm}
\end{figure}

By definition, any diffractive pattern in ${d\sigma}$/$dt$ is a
property of the (predominantly imaginary) Pomeron contribution,
rather than of the ``non-diffractive" Odderon, whose amplitude is
predominantly real and thus it can only ``contaminate" the
diffractive pattern.

\subsection{A simple model of the diffraction pattern}

Data on $pp$ scattering below the LHC energy region can be
described by a Pomeron and sub-leading contributions. From the
phenomenological point of view, the need for the Odderon, becomes
important only after the SPS data on $p\bar p$ scattering
appeared, in which the dip seen in $pp$ scattering appeared as a
``shoulder", suggesting that the diffractive minimum was filled by
the Odderon contribution. This is not a proof, just a plausibility
argument. Below we illustrate the dynamics of the dip and shoulder
in $pp$ and $p\bar p$ scattering by using a simple (``minimal")
dipole Pomeron (DP) model to ``guide the eye". It reproduces the
observed dynamics of the dip and will help to anticipate the LHC
phenomena.

By neglecting spin effects, the scattering amplitude for $pp$ and
$p\bar p$ scattering is written as a sum
\begin{eqnarray}
    A^{p\bar p}_{pp}(s,t)=P(s,t)\pm O(s,t), \label{PO}
\end{eqnarray}
where $P$ is the Pomeron contribution and $O$ is that of the
Odderon. For the DP amplitude we use the following ``minimal"
model \cite{Shelk_a, Shelk_b, Vall, Covolan_a, Covolan_b}

\begin{eqnarray}
  P(s,t)=i{as\over{bs_0}}\Bigl(r^2_1(s)e^{r^2_1(s)[\alpha_P(t)-1]}-
\epsilon r^2_2(s)e^{r^2_2(s)[\alpha_P(t)-1]}\Bigr), \label{DP}
\end{eqnarray}
where $r_1^2(s)=b+L-{i\pi\over 2},\ \ r_2^2(s)=L-{i\pi\over 2}$
with $L\equiv\ln{s\over{s_0}}$; $ \alpha_P(t)$ is the Pomeron
trajectory and $a,b,s_0$ and $\epsilon$ are free parameters. The
model produces rising cross sections without violation of the
Froissart bound as well as a dip-bump structure seen in the $pp$
scattering with its non-trivial dynamics observed at the CERN ISR
in the range $23<\sqrt s<62 GeV $ \cite{Shelk_a, Shelk_b, Vall,
Covolan_a, Covolan_b}. The absence of a relevant structure
(replaced by a ``shoulder") in $p\bar p$ scattering at the CERN
SPS was interpreted \cite{Shelk_a, Shelk_b, Covolan_a, Covolan_b}
as a manifestation of the Odderon, filling the dip produced by the
Pomeron term (for reviews see Refs.~\cite{Jenk, VJS1}). A
high-quality description of elastic high-energy data with
predictions for future accelerators can be found in
\cite{Predazzi}. In an alternative successful approach
\cite{Prokudin} there are several Pomerons, whose interference
produces the dip-bump structure and perfect agreement with the
data.

Since little is known about the properties of the Odderon, apart
from its assumed asymptotic nature, one usually parameterizes it
in the form close to that of the Pomeron given in Eq.~(\ref{DP}).
However: 1)~the Odderon is $C-$ odd, which implies an extra $i$
factor in front of the amplitude; 2)~its relative contribution is
by orders of magnitude smaller than that of the Pomeron (until now
the Odderon was not seen in the forward direction); 3)~the slope
of the Odderon trajectory is much smaller than that of the
Pomeron. There are two reasons for the latter: one is based on
theoretical arguments \cite{Shelk_a,Shelk_b} and the other one is
phenomenological: the ``flat'' shoulder in $p\bar p$ scattering,
seen at the SPS, could be a manifestation of the Odderon.

Let us remind that the dip in the ISR energy region is
monotonically deepening, reaching its maximal depth at $52.8$ GeV,
whereafter the monotonic trend changes, albeit at a single energy
equal to $62$ GeV. Similar to the unique case of the measured
\cite{dip_exp_a,dip_exp_b} difference between $pp$ and $p\bar p$
amplitudes at $53$ GeV, mentioned in the Introduction, this
phenomenon needs confirmation. Regrettably, measurements of the
difference between $pp$ and $p\bar p$ amplitudes are not foreseen
in the near future.


\begin{figure}[tbh!]
\includegraphics[width=0.5\textwidth]{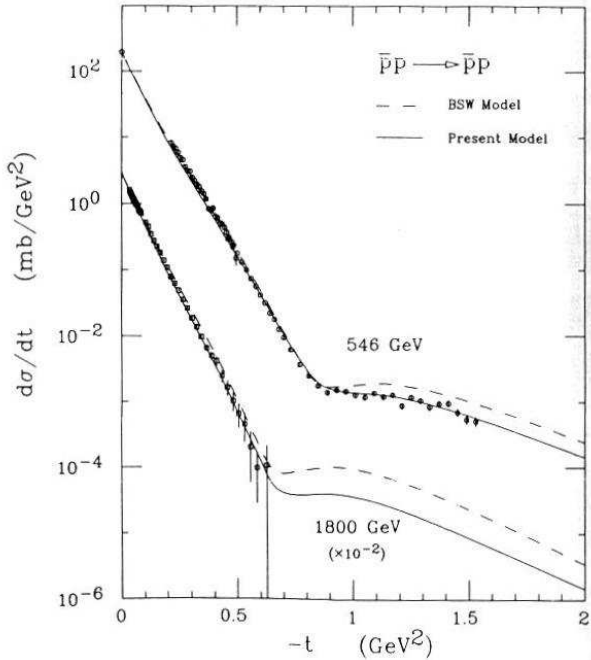}
\caption{Differential cross section at $546$ and $1800$ GeV
calculated from Eqs.~(\ref{PO}) and (\ref{DP}) with the $P+O$
contribution, corresponding to $p\bar p$ scattering (full line),
compared with those of Ref. \cite{BSW} (broken line). For detail
of the calculations see Refs. \cite{Shelk_a, Shelk_b, Covolan_a,
Covolan_b, Vall}.} \vspace{0.5cm} \label{Dip11.eps}
\end{figure}

Fig. \ref{Dip11.eps} shows the differential cross section of
$p\bar b$ scattering at $546$ and $1800$ GeV calculated from
Eqs.~(\ref{PO}) and (\ref{DP}) (full line), compared with a fit
from paper \cite{BSW} (broken line). Details of the calculations
and the values of the fitted parameters can be found in Refs.
\cite{Shelk_a, Shelk_b, Covolan_a, Covolan_b, Vall}.


It should be reminded that the hypothetical Odderon contributes
enters Eq.~(6) with different signs in $pp$ and $p\bar p$
amplitudes, thus distorting the pure Pomeron contribution in both
cases. Due to its small slope (flatness of its trajectory), the
role of the Odderon increases with increasing $-t,$ the phenomenon
reaching its maximum in the dip region.

 The ratio of the cross section at the minimum
to that at the maximum (depth of the dip) is more informative than
its absolute value. Even greater than at the ISR ($\sim 5$) value
of the ratio will indicate the monotonic deepening of the dip and
will disfavour the Odderon contribution, while its shallowing will
favour the presence of the Odderon.

At this point it may be appropriate to introduce several known
models describing the diffractive pattern (the dynamics of the
dip-bump structure). Before doing so, two points are worth
mentioning: 1) one still lacks a theoretical understanding of the
phenomenon and must rely on models; 2) among the large number of
models, only few are able to fit properly the large number of
high-statistics data that exist in a wide range of $s$ and $t$ for
$pp$ and $\bar pp$ scattering. A collection/review of the existing
models for high-energy scattering with a critical evaluation of
their prediction is highly desirable, however this task is beyond
the scope of the present paper. We apologize to those authors
whose papers, for brevity of space, are not cited here.

\subsection{The ``Protvino'' model
\label{sec:protvino}}
One of the examples of eikonal models of diffraction is
the ``Protvino'' model \cite{Prokudin03,Prokudin}.

 The unitarity condition
$$
\Im{\rm m}\; T(s,\vec b) = \vert T(s,\vec b)\vert^2 + \eta (s,\vec b)\; ,
$$
where $T(s,\vec b)$ is the scattering amplitude in the impact
representation,
$\vec b$ is the impact parameter, $\eta (s,\vec b)$ is the
contribution of
inelastic channels, implies the following eikonal form for the scattering amplitude $T(s,\vec b)$:
\begin{equation}
T(s,\vec b)=\frac{e^{2i\delta (s,\vec b)}-1}{2i}\; ,
\label{eq:ampl}
\end{equation}
where $\delta (s,\vec b)$ is the
eikonal function. The unitarity condition in terms of the eikonal looks as follows:
\begin{equation}
{\rm \Im m}\; \delta (s,\vec b) \ge 0, \; s>s_{\rm inel}\; .
\label{eq:euc}
\end{equation}

The eikonal function is assumed to have simple poles in the
complex $J$-plane and the corresponding Regge trajectories are
normally used in the linear approximation \be \alpha(t) =
\alpha(0) + \alpha '(0)t\; . \label{eq:rt} \ee

The following representation for the eikonal function is used:
\begin{equation}
\delta_{pp}^{\bar p p}(s,b) = \delta^+_{{P}_1}(s,b)+
\delta^+_{{P}_2}(s,b)+
\delta^+_{{P}_3}(s,b)
\mp \delta^-_{O}(s,b)+\delta^+_{
f}(s,b)\mp \delta^-_{\omega}(s,b),
\label{eq:modeleik}
\end{equation}
where $\delta^+_{{P}_{1,2,3}}(s,b)$ are Pomeron contributions.
`$+$' denotes C even trajectories (the Pomeron trajectories have
the following quantum numbers $0^+J^{++}$), `$-$' denotes  C odd
trajectories, $\delta^-_{O}(s,b)$ is the Odderon contribution (the
Odderon is the C odd partner of the Pomeron with quantum numbers
$0^-J^{--}$); $\delta^+_{ f}$, $\delta^-_{\omega}(s,b)$ are the
contributions of secondary Reggeons, $f$ ($C=+1$) and $\omega$
($C=-1$).

The parameters of secondary Reggeon trajectories are fixed according to
the parameters obtained from a fit of the meson spectrum:
\be
\begin{array}{l}
\alpha_f(t) = 0.69+0.84 t\; , \\
\\
\alpha_\omega (t) = 0.47+0.93 t\; .
\end{array}
\ee

\begin{figure}[t]
\includegraphics[width=0.5\textwidth,bb= 10 140 540 660,angle=0]
{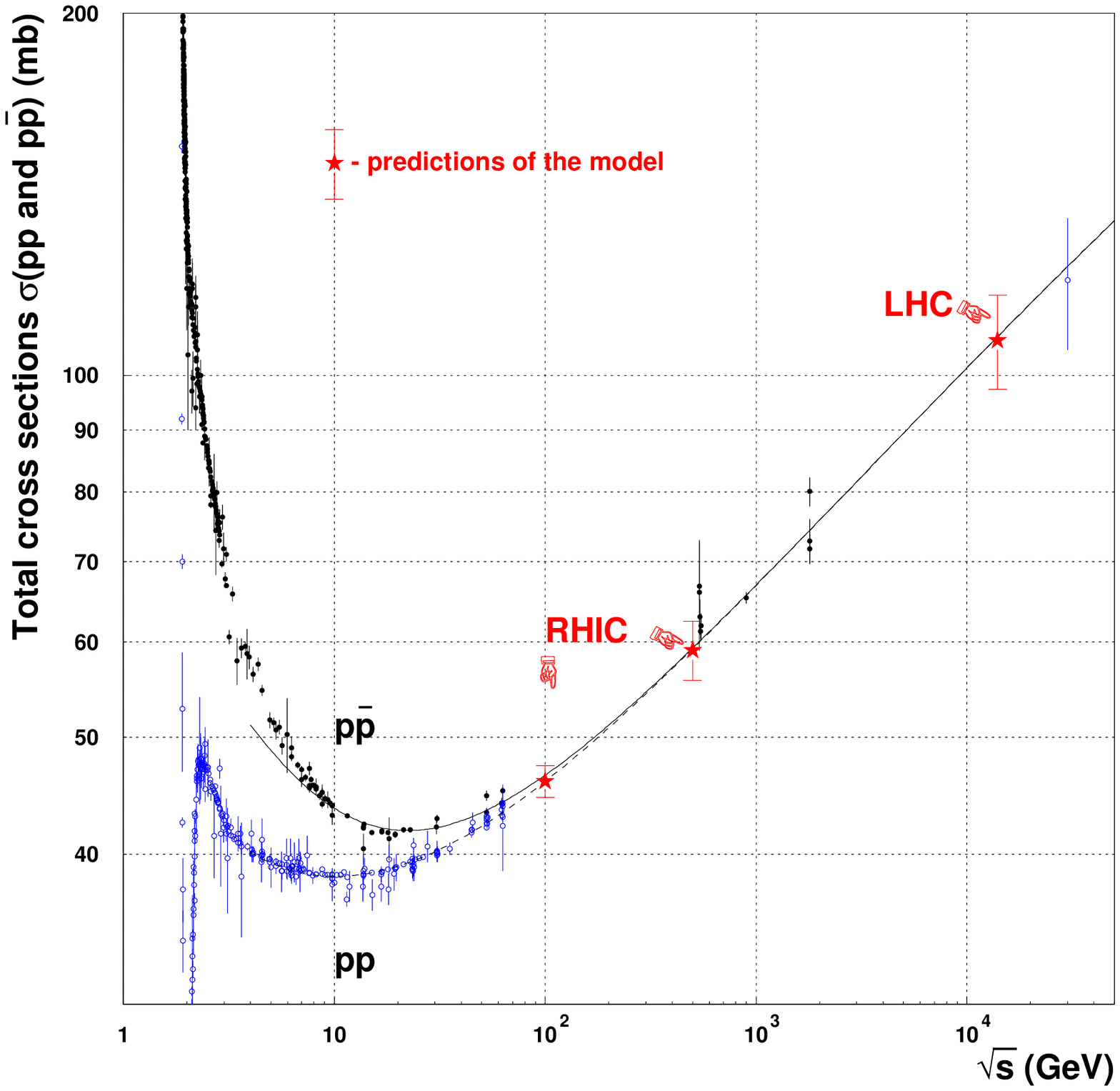}
\includegraphics[width=0.5\textwidth,bb= 10 140 540 660,angle=0]
{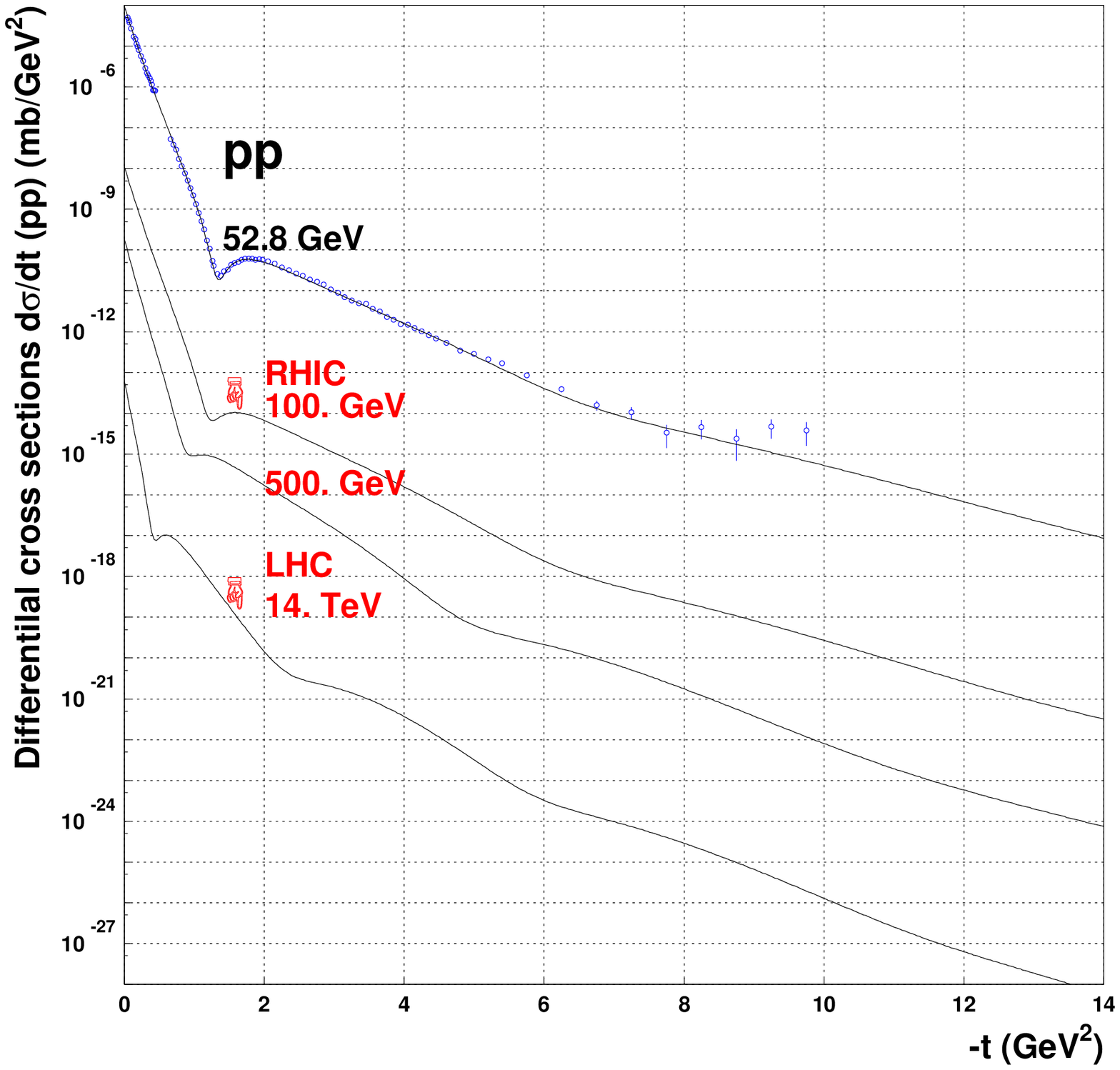} \caption{ Left panel: Total cross sections of $pp$
scattering (hollow circles)  and $\bar p p$ scattering (full
circles) and curves corresponding to their description in the
three-Pomeron model  \cite{Prokudin, Prokudin03}. Right panel:
Predictions of the three-Pomeron model  \cite{Prokudin,
Prokudin03} for the differential cross-section of $p p$ scattering
to be measured at LHC with $\sqrt{s}=14.\; TeV$ and at RHIC
$\sqrt{s}=100.\; GeV$ and $\sqrt{s}=500.\; GeV$. The data
corresponding to the energy $\sqrt{s}=52.8\; GeV$ is multiplied by
 $10^{-6}$, RHIC at $500 \ $GeV by $10^{-10}$, RHIC at $500 \ $GeV by $10^{-12}$,
 and that of LHC by $10^{-16}$.
\label{fig:tot}}
\end{figure}

The model fits high-energy elastic $pp$ and $\bar p p$ scattering
data. The data are well described for all momenta ($0.01\le |t|\le
14.\; GeV^2$) and energies ($8.\le\sqrt{s}\le 1800 \ $GeV)
($\chi^2/{\rm d.o.f.}=2.60$). It predicts the appearance of two
dips in the differential cross-section which will be measured at
LHC. The parameters of the Pomeron and Odderon trajectories are:
 \\
$\alpha(0)_{{P}_1}=1.058,\;\;\alpha'(0)_{{P}_1}=0.560\; GeV^{-2};$ \\
$\alpha(0)_{{P}_2}=1.167,\;\;\alpha'(0)_{{P}_2}=0.273\; GeV^{-2};$ \\
$\alpha(0)_{{P}_3}=1.203,\;\;\alpha'(0)_{{P}_3}=0.094\; GeV^{-2}.$ \\
$\alpha(0)_{{O}}=1.192,\;\;\alpha'(0)_{{O}}=0.048\; GeV^{-2}.$ \\

The model predicts (see Fig.~\ref{fig:tot}) the following values
of total and elastic cross sections at the LHC:
\begin{eqnarray}
\nonumber
\sqrt{s}=14.\; TeV \;, \\
\nonumber
\\
\sigma_{tot}^{pp} = 106.73\;\;(mb)\;_{-\; 8.50mb}^{+7.56 \;mb}\;, \\
\nonumber
\\
\nonumber
\sigma_{elastic}^{pp} = 29.19\;\;(mb)\;_{-2.83\; mb}^{+3.58 \;mb}\;, \\
\nonumber
\\
\nonumber
\rho^{pp} = 0.1378\;_{-0.0061}^{+0.0042}\;.
\end{eqnarray}

On the right panel of Fig.~\ref{fig:tot} one can see a typical
``diffractive'' pattern of the differential cross-section also
present in other models based on unitarisation~\cite{Predazzi66,
Predazzi79_a, Predazzi79_b, Soffer_a, Soffer_b, SelyuginYF80} (see
also Fig.~\ref{Dip11.eps} and Fig.~\ref{fig:islam_tot}).

\subsection{The ``Connecticut'' model \label{sec:islam}}

A model based on a physical picture of the nucleon having an
external cloud, an inner shell of baryonic charge, and a central
quark-bag containing the valence quarks was proposed in
Ref.~\cite{Islam}. The underlying field theory model is the gauged
Gell--Mann -- Levy linear $\sigma$-model with spontaneous breakdown
of chiral symmetry, with a Wess--Zumino--Witten (WZW) anomalous
action. The model attributes the external nucleon cloud to a
quark--antiquark condensed ground state analogous to the BCS
ground state in superconductivity -- an idea that was first
proposed by Nambu and Jona-Lasinio.  The WZW action implies that
the baryonic charge is geometrical or topological in nature, which
is the basis of the Skyrmion model.  The action further shows that
the vector meson $\mathrm{\omega}$ couples to this topological
charge like a gauge boson, i.e.\ like an elementary vector meson.
As a consequence, one nucleon probes the baryonic charge of the
other one via $\mathrm{\omega}$-exchange.  In pp elastic
scattering, in the small momentum transfer region, the outer cloud
of one nucleon interacts with that of the other giving rise to
diffraction scattering.  As the momentum transfer increases, one
nucleon probes the other one at intermediate distances and the
$\mathrm{\omega}$-exchange becomes dominant.  At momentun
transfers even larger, one nucleon scatters off the other via
valence quark-quark scattering.

Diffraction is described by using the impact parameter
 representation and a phenomenological
profile function:\vspace{-0.2cm}

\begin{figure}[t]
\includegraphics[width=0.5\textwidth,bb= 10 140 540 660,angle=0]
{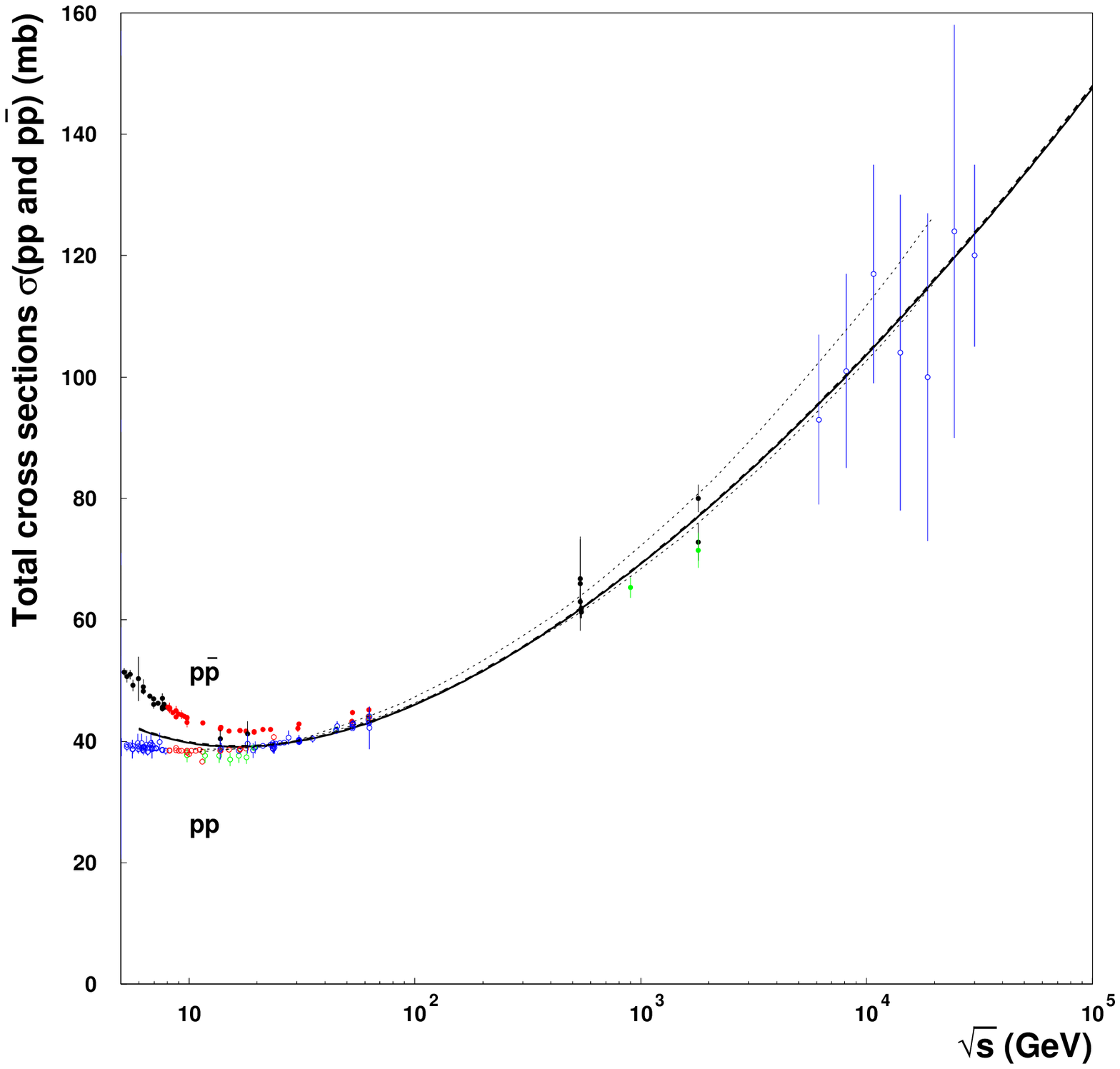}
\includegraphics[width=0.5\textwidth,bb= 10 140 540 660,angle=0]
{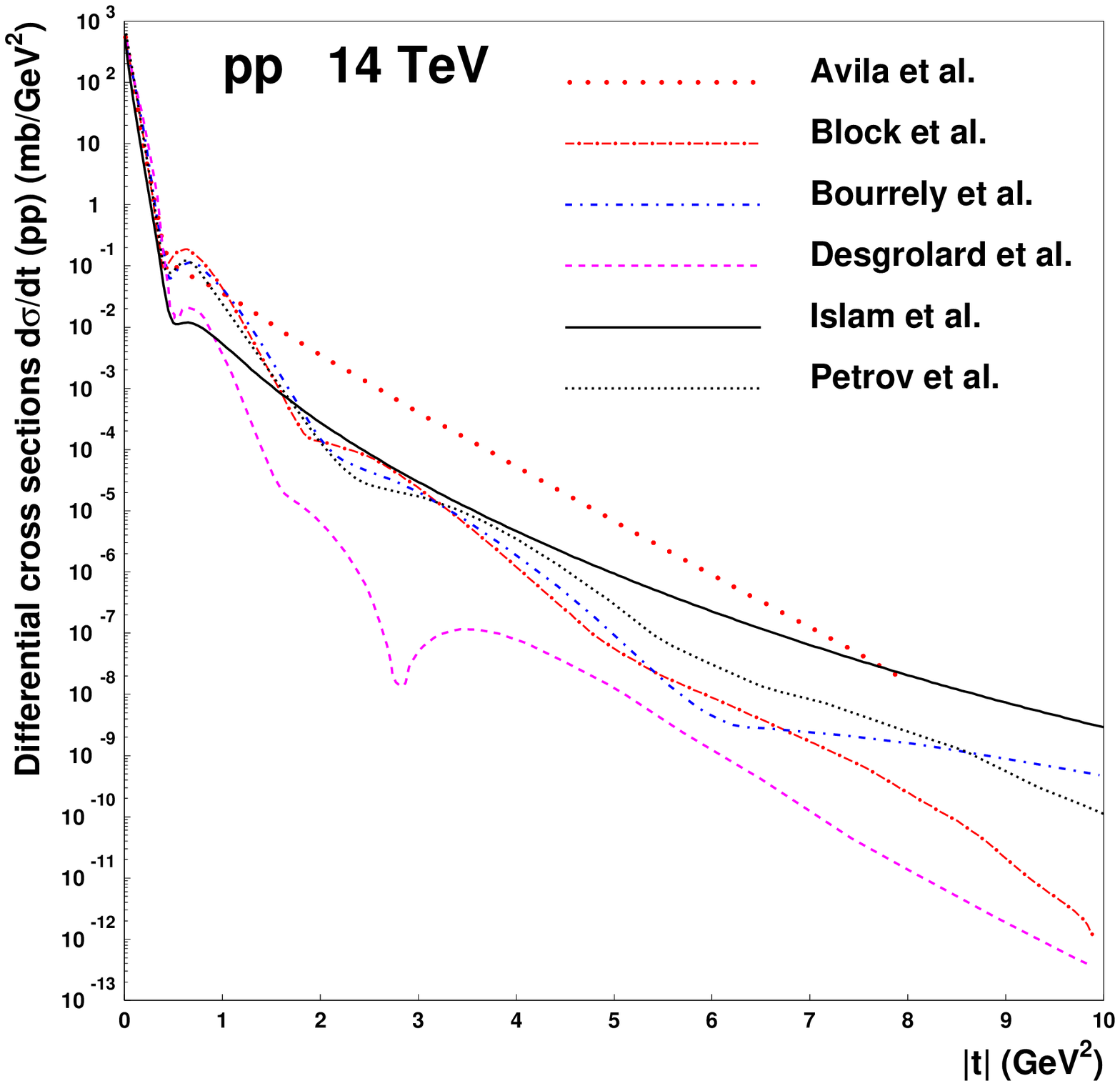} \caption{ Left panel: Total cross sections of $pp$
scattering (open circles)  and $\bar p p$ scattering (full
circles) and the fitted curves based on the model \cite{Islam}.
Right panel: Predictions of the model \cite{Islam} for the
differential cross-section of $p p$ scattering to be measured at
the LHC at $\sqrt{s}=14.\; TeV$ (solid line) and comparison with
the prediction of other models: Avila et al.\ \cite{Nicolescu},
Block et al.\ \cite{Block}, Bourrely et al.\
\cite{Soffer_a,Soffer_b}, Desgrolard et al.\
\cite{GiffonMartynov}, and Petrov et al.\ (three pomerons)\
\cite{Prokudin03}. \label{fig:islam_tot}}
\end{figure}

\begin{equation}
T_D (s,t)=i\,p\,W\int_0^\infty {b\;\rm{d}\it{b}\;J_0 } (b\,q)\Gamma_{D}(s,b)\;.
\end{equation}
here $q$ is the momentum transfer ($q = \sqrt{|t|}$) and
$\Gamma_{D}(s,b)$ is the diffraction profile function, which is
related to the eikonal function $\chi_{D}(s,b)$: $\Gamma_{D}(s,b)
= 1 - \exp(i \chi_{D}(s,b))$. $\Gamma_{D}(s,b)$ is taken to be an
even Fermi profile function:

\begin{equation}
\Gamma_{D}(s,b) = g(s) \left[\frac{1}{1 + e^{(b-R)/a}} +  \frac{1}{1 + e^{-(b+R)/a}} -1 \right]\; .
\end{equation}

The parameters $R$ and $a$ are energy dependent:
$$R  =  R_0 + R_1\left(\ln \frac{s}{s_0} -  i\frac{\pi}{2}\right),$$
$$a = a_0 + a_1\left(\ln \frac{s}{s_0} - i\frac{\pi}{2}\right);$$ $g(s)$ is a complex crossing even
energy-dependent coupling strength, $s_0 = 1$ GeV$^2$.

For the diffraction amplitude the following asymptotic properties are obtained:
\begin{enumerate}
\item   $\sigma_{\rm{tot}}(s) \sim (a_{0} + a_{1} \ln s)^{2}\hspace{2.82cm}$ (Froissart-Martin bound)\vspace{0.1cm}\\
\item   $\rho (s) \simeq \frac{\pi a_1}{a_0 + a_1 \ln s}\hspace{4.13cm}$ (derivative dispersion relation)\vspace{0.06cm}\\
\item   $T_{D}(s,t) \sim i\; s\; \ln^{2}s\; f(|t|\; \ln^{2}s)\hspace{1.72cm}$  (Auberson-Kinoshita-Martin scaling)\vspace{0.1cm}\\
\item   $T_{D}^{\mathrm{\bar{p}p}}(s,t) = T_{D}^{\mathrm{pp}}(s,t)\hspace{3.39cm}$      (crossing even)
\end{enumerate}

At 14 TeV $\sigma_{tot}$ and $\rho_{pp}$ were found to have the
values 110 mb and 0.120 mb, respectively.

Finally, Fig.~\ref{fig:islam_tot} shows a compilation of
predictions of the model for pp elastic $\rm{d}\sigma/d\it{t}$ at
LHC together with the predictions of other models, e.g.: Avila et
al.\ \cite{Nicolescu}, Block et al.\ \cite{Block}, Bourrely et
al.\ \cite{Soffer_a,Soffer_b}, Desgrolard et al.\
\cite{GiffonMartynov}, and Petrov et al.\ (three pomeron)\
\cite{Prokudin03}. As one can see from Fig.~\ref{fig:islam_tot},
the model \cite{Islam} predicts no diffractive pattern at large
$|t|$ as diffractive scattering dominates in the small $|t|$
region, hard scattering dominates in the intermediate $|t|$ region
($1.5 <  |t| < 8.0\;{\rm GeV^2}$) and quark-quark scattering takes
over at large $|t|$.

\subsection{The ``Dubna Dynamical'' model \label{sec:ddmodel}}

The so-called ``Dubna Dynamical'' (DD) model was already cited in
connection with the nearly forward scattering; it will be used
also in Sec. 6.2 on the BDL, so we find it appropriate to briefly
introduce it, especially as it may be closely related to models
proposed earlier (see
Refs.~\cite{Soffer_a,Soffer_b,Predazzi66,Predazzi79_a,Predazzi79_b})
or later (see Refs.~\cite{ Islam, Prokudin, Prokudin03} and
earlier references therein).

The DD model was developed in
Refs.~\cite{SelyuginYF80,SelyuginYF82_a, SelyuginYF82_b} to
describe particle interaction with account for the internal
structure of hadrons. According to the DD model, the hadron
scattering amplitude can be written as a
  sum  of a central and a peripheral part of the interaction
  \cite{SelyuginYF80}.  
The full scattering amplitude is calculated using the impact
 parameter representation with the eikonal form of the unitarization:
\begin{equation}
 A(s,t) = is \int_{0}^{\infty} \ b db J_{0}(b\Delta)
 [1-e^{-\chi(s,b)}],
\end{equation}
where
\begin{equation}
 \chi(s,b) = \chi_{c}(s,b) + \chi_{p}(s,b)+\chi_{R}(s,b).  \label{tt}
\end{equation}
Here the central interaction term $\chi_{c}(s,b)$  describes  the
interaction between central parts of the hadrons. At high
energies, it is determined by a spinless Pomeron exchange. The
second term, $\chi_{p}(s,b)$ is the sum of triangle diagrams
corresponding to the interaction of the central part of a hadron
with the  meson cloud of the other one. The meson-nucleon
interaction leads to spin flip effects in the Pomeron-hadron
vertex \cite{SelyuginNP87}. The term $\chi_{R}(s,b)$
describe the cross even and cross odd parts of the subleading
Reggeons with the energy dependence $\sim 1/\sqrt{s}$. 

\begin{figure}[tbh!]
\begin{center}
\includegraphics[width=0.7\textwidth]{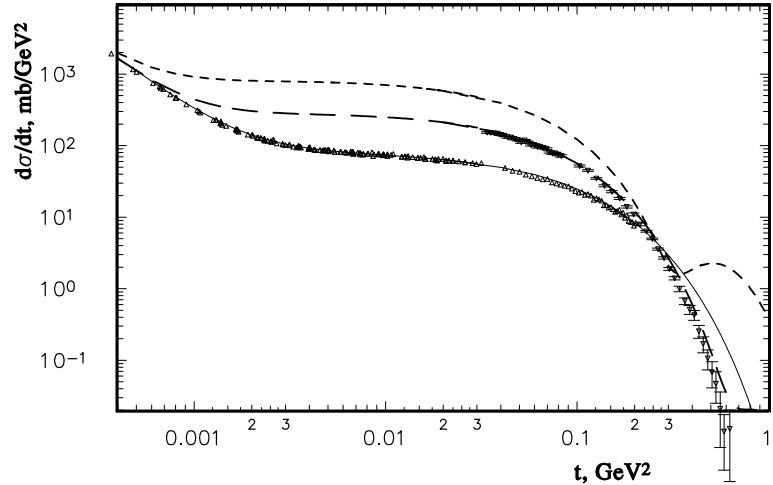}
\end{center}
\caption{Predictions of the DD model
\cite{SelyuginPN87_a,SelyuginPN87_b} for the Tevatron energies,
$\sqrt{s}=1.8$ TeV (long dashed line) and for the LHC energies
(short dashed line). For the sake of comparison, model curves and
experimental data are shown for one lower energy, $\sqrt{s}=23$
GeV (solid line). The experimental data points are from
Ref.~\cite{ss23}.} \vspace{0.5cm} \label{prediction-DDm}
\end{figure}


The term $\chi_{c}(s,b$ ) was chosen phenomenologically (for
details see Refs.~\cite{Predazzi66,Predazzi79_a,Predazzi79_b}),
where the scattering amplitude was obtained from an integral
representation of the Bessel function. It was shown in those
papers that, starting from the usual partial wave expansion in the
Legendre polynomials, the eikonal representation for the
scattering amplitude can be derived to be valid for all energies
and in the whole angular domain, and for which the spectral
function is given in terms of the partial wave amplitudes. The
scattering amplitude can be presented in the form
\begin{equation}
A(p,\Theta)\approx
-iB\alpha^{-\nu}\beta^{1-\nu}p{K_{\nu-1}[\beta(\alpha^2+p^2\sin
^2\theta)^{1/2}]\over{\alpha^2+p^2\sin^2\Theta^{1-\nu}/2}}.
\end{equation}
Although the analytic calculations can be performed to any $\nu$,
to simplify numerical calculations, we set $\nu=1/2$, whereafter
the amplitude becomes:
\begin{equation}
A(p,\Theta)=
-i(\pi/2)^{1/2}\alpha^{-1/2}Bp{\exp[-\beta(\alpha^2+p^2\sin
^2\theta)^{1/2}]\over{\alpha^2+p^2\sin^2\Theta^{1-\nu}/2}}.
\end{equation}
  By taking this form as the Born amplitude, the eikonal term  $\chi_0$ in the DD
model assumes the form
  \begin{equation}
\chi_{0}(s,b)= A_0 \ e^{-m(s) \ \sqrt{R_{0}^2 + b^2}}.
\end{equation}
 It leads to an exponential behavior of the eikonal
 at large values of the impact parameter and a Gaussian
behavior at small impact parameters.
  In the DD model the inelastic states in the $s$-channel were also taken
into account.
  Accordingly, the central part can be presented as
\begin{equation}
\chi_{c}(s,b) =  \chi_{0}(s,b) \ [1- \gamma \chi_{0}(s,b)],
\end{equation}
with
\begin{equation}
\chi_{0}(s,b)= A_0 \ e^{-m(s) \ \sqrt{R_{0}^2 + b^2}}.
\end{equation}
 The coefficient $\gamma$ is connected with the contribution of the inelastic
 intermediate state in the $s$-channel. Here $A_0,\ m(s)$ and $\ R_0$
 are free parameters, whose values and energy dependence were determined
 from fits to the experimental data on proton-proton elastic differential cross
 section scattering from $\sqrt{s} = 9.78 \ $GeV
 to  $\sqrt{s} = 540 \ $GeV (see Fig.~\ref{prediction-DDm}) and momentum transfer from $t=0$
 to $t= -12 \ $GeV$^2$.

The additional term $\chi_{p}$ was determined by the peripheral
Pomeron interaction with account for the meson cloud. This leads
to the important feature of the DD model, namely the existence in
the eikonal phase of a small term growing as $\sqrt{s}:$
\begin{equation}
\chi_p(s,b) = H_{p0} + \lambda \sqrt{s} \ m_{p}\sqrt{R_{p}^2+b^2}
K_{1}(\ m_{p}\sqrt{R_{p}^2+b^2}),
\end{equation}
where $K_{1}$ is the MacDonald function of the first order.
So, the contribution to the eikonal phase growing as $\sqrt s$ is
determined by the peripheral meson-cloud effects. It has been
shown that this term becomes important for energies $\sqrt{ s}
> 200 \ $ GeV.
  The corresponding form of the interaction potential can be
  represented, see Ref.~\cite{Predazzi66}, by a superposition of
  Yukawa-like potentials in the form of the MacDonald function.
  The leading term of the $\chi_{c}$ in the momentum transfer representation
  is
\begin{equation}
 \sim  {\ e^{-R_{0} \ \sqrt{4 m_{p}^2 + q^2}}\over{ \sqrt{4
m_{p}^2+ q^2}}}.
\end{equation}
  It results in the following form of the differential cross section
for $q^2 >> m_{p}^2$
  \cite{SelyuginYF80}
\begin{equation}
 \frac{d \sigma}{dt}  \sim \frac{ \ e^{-2 R_{0} \sqrt{|t|} }}{t^2}.
\end{equation}

Notice that the square-root behavior in the above formula
corresponds to the square-root Pomeron trajectory, Eq. (2), both
following from analyticity and unitarity, with important physical
consequences, such as the ``break" in the cone near $t=-0.1$
GeV$^2$ \cite{Jenk, Gribov_a} or levelling off of the impact
parameter amplitude at large $b$, corresponding to the mesonic
atmosphere of the nucleon.

        The model provides  a self-consistent  picture  of
the differential cross section and spin phenomena of different
hadron processes  at high  energies. Really, the parameters in the
amplitude  determined from one reaction, for example, elastic
$pp$-scattering, allow to give predictions on elastic
meson-nucleon scattering and charge-exchange reaction
 $\pi^{-} p \rightarrow  \pi^{0} n$
 at high energies.
     The model predicts that at super-high energies polarization
effects for particles and antiparticles are the same
\cite{SelyuginNP87,SelyuginZC91}.

\begin{figure}[tbh!]
\vspace{-2cm}
\begin{center}
\includegraphics[width=0.9\textwidth]{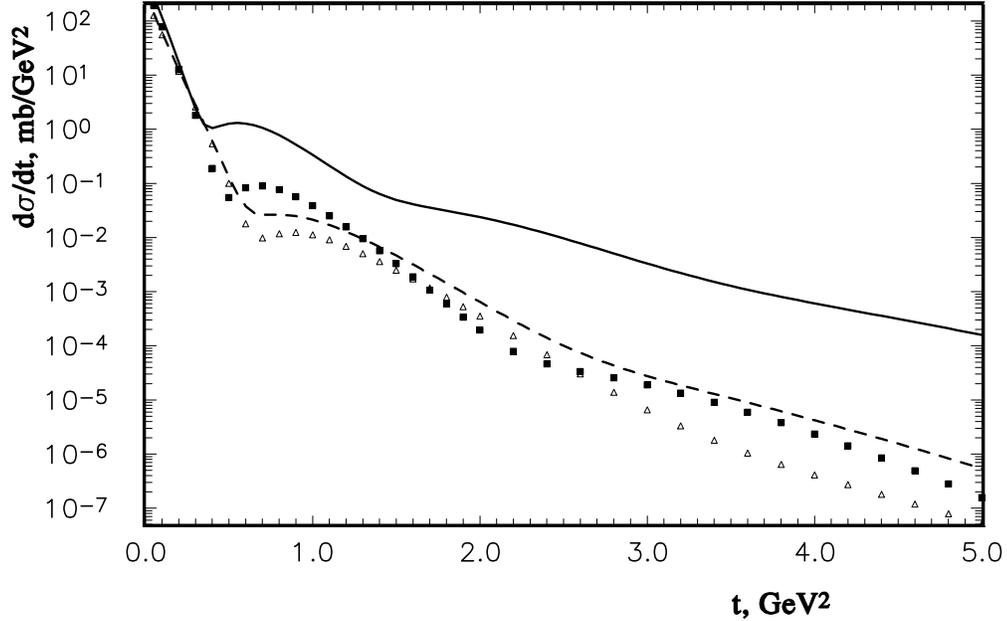}
\end{center}
\caption{Predictions of the DD (solid and dashed lines) and BSW
models (triangles and full squares) for the differential cross
section,  at $\sqrt{s} = 2 \ $TeV (dashed line and triangles) and
$\sqrt{s}=10 \ $ TeV (solid line and full squares).
} \vspace{0.5cm}
\label{fig:dsppmod.ps}
\end{figure}


 Let us note that a similar model, describing simultaneously
 different
  hadronic elastic cross sections, polarization and spin correlation
  effects was developed by Bourrely, Soffer and Wu \cite{Soffer_a,Soffer_b} (BSW).
  According to the physical picture of this model, each hadron
  appears as a black disk with a gray fringe, where the black disk
  radius increases as $\ln (s)$.
    It is interesting to compare the predictions of the
 BSW  and  DD models for the LHC energies. The result is shown in
 Fig.~\ref{fig:dsppmod.ps} and Table \ref{table:dd_BSW}.
    The most important difference comes from the region $|t| > 0.3 $ GeV$^2$.

\begin{table*}
\begin{tabular}{lllllllll}
\hline\noalign{\smallskip}
  energy  &    model &   $\sigma_{tot}$, mb & $ \sigma_{el}$, mb
&$ \frac{\sigma_{el}}{\sigma_{tot}}$ & $\rho(t=0)$ &$ \frac{d
\sigma}{dt}|_{-t=1} \ $mb/GeV$^2$
             \\
\hline\noalign{\smallskip}
   2 TeV & DD & 81 &   20.7  &  0.256 83 &  0.197  & 38.8 \\
     & BSW & 76.1 &  17.9  & 0.235 82 &    0.128 &  11.2 \\
\hline\noalign{\smallskip}
   10 TeV  & DD & 123 &   42.6   & 0.35 &    0.195  &  350 \\
     & BSW &  98.4 &   26.8   & 0.27       &   0.122 & 38.6 \\
\noalign{\smallskip}\hline
\end{tabular}
\caption{Comparison of the predictions from the DD and BSW models
for two LHC energies. \label{table:dd_BSW}}
\end{table*}

Future experiments at the LHC will provide for an excellent
possibility to
 test the theoretical argument on the peripheral rise of the
eikonal phase at superhigh energies determined, in this approach,
by the meson-cloud effects.

\section{Intermediate and Large $|t|$} \label{s5}

The definition of the transition between ``soft" and ``hard"
physics is rather ambiguous, although there are phenomena
indicative of this transition. One is the slow-down of the
shrinkage of the cone, already noticed at the ISR, and attributed
to the hard scattering of the constituents of nucleons. It will
manifest also in the change of the exponential decrease to a power
fall-off in $|t|$. Furthermore, in this region the exponential
decrease of the scattering amplitude will be gradually replaced by
a power one. These effects have a simple interpretation within a
Regge-pole model with nonlinear trajectories. As shown in a number
of papers (see e.g. Refs.~\cite{scaling, Fioreetal}), logarithmic
trajectories in the dual Regge-model \cite{JenkNP, DAMA_c} can
mimic hard scattering of point-like constituents and the
transition from the soft interaction of the extended objects, like
strings, to the hard scattering of point-like particles, governed
by the quark counting rules and/or QCD.


\begin{figure}[tbh!]
\begin{center}
\includegraphics[width=0.7\textwidth]{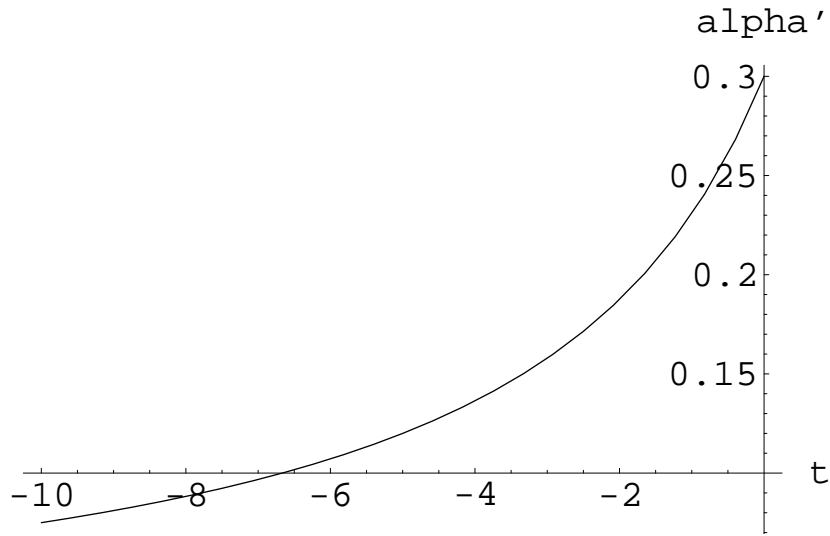}
\end{center}
\caption{Typical behavior of the slope of a logarithmic
trajectory. Using Eq.~(\ref{log1}), the slope $\alpha'(t)$
decreases between $t=-1$ and $t=-10$ by about a factor 3.
According to Eq.~(\ref{slope}) the shrinkage of the slope is
expected to slow down by the same rate.} \vspace{0.5cm}
\label{fig:SlopeLHC.eps}
\end{figure}


A particularly simple expression for the slope of the diffraction
cone, illustrating the aforesaid phenomena, follows from the
geometrical properties of the geometrical model (see Eq.~\ref{DP}
and Ref.~\cite{VJS}):
\begin{eqnarray}
B(s,t)=k\alpha'(t) \sigma_t(s),
\label{slope}
\end{eqnarray}
where $k$ is a parameter and $\alpha'(t)$ is the slope of the
logarithmic Pomeron trajectory given by Eq.~(\ref{log}). By
choosing $\gamma=1$ and $\beta=0.3$, we get from Eq.~(\ref{log})

\begin{eqnarray}
\alpha'(t)=\frac{0.3}{1-0.3 t}.
\label{log1}
\end{eqnarray}

The slope $\alpha'(t)$ given by Eq.~(\ref{log1}) is shown in
Fig.~\ref{fig:SlopeLHC.eps}. The logarithmic trajectory
(\ref{log}) (or, similarly, (\ref{combined})) mimics also the
transition from the exponential decrease of the scattering
amplitude in $t$ (soft physics) to a power one, according to

\begin{eqnarray}
 A(s,t)\sim e^{[b\alpha(t)]}s^{\alpha(t)}\sim e^{-[b+\ln(s)]\ln(-t)}. \label{nl-tr}
\end{eqnarray}


\begin{figure}[tbh!]
\begin{center}
\includegraphics[width=0.9\textwidth]{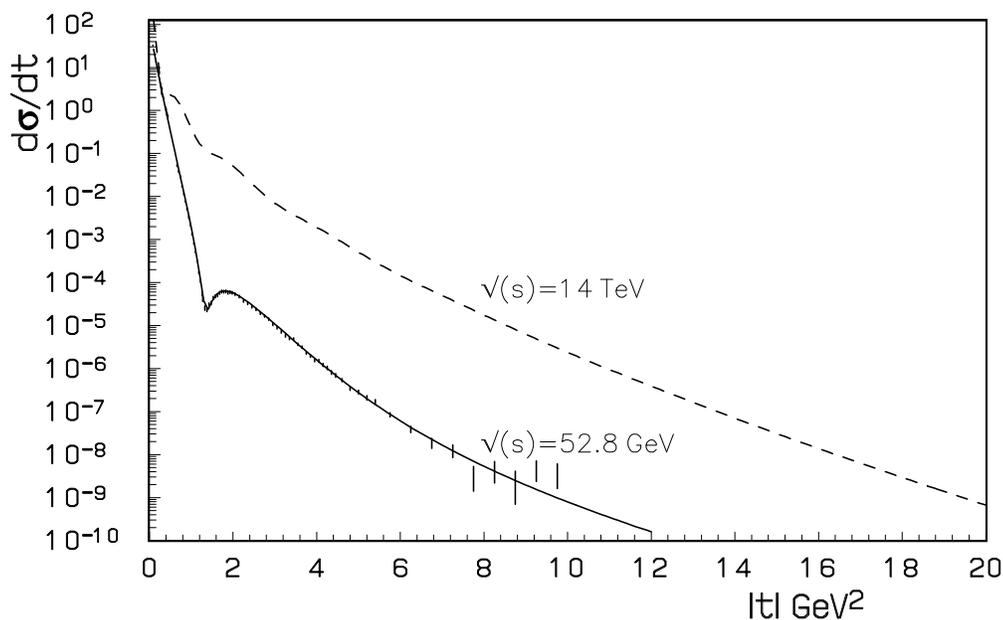}
\end{center}
\caption{Predictions of the DD model for the differential cross
section at $\sqrt{s} = 52 \ $GeV (solid line) and at $\sqrt{s} =
14 \ $TeV (dashed line) \cite{SelyuginPN87_a,SelyuginPN87_b}}.
\label{fig:ds14tev.ps} \vspace{0.5cm}
\end{figure}


Note that in this region the ratio $|t|/s$ is much smaller (tends
to zero) than at the ISR energies. The new information obtained in
this region (small angles and large $t$)
 will give the possibility to determine whether perturbative QCD
works here. The logarithmic regime of the trajectory in the above
Regge-pole model with the resulting power-like behavior in this
kinematical region provides a link with the quark-parton model
and, eventually, with the perturbative regime of QCD. However this
transition is much more complicated, involves a rearrangement of
the dynamics, that can be treated e.g. in the dual model of Ref.
\cite{Fioreetal}.

The valence $qq$ scattering is assumed to be due to a color
singlet amplitude behaving as $s/t$, so that $d\sigma_{qq}/dt \sim
t^{-2}$. For large $|t|$, this leads to a power-like behavior
\cite{s-s}: $d\sigma/dt \sim t^{-10}$, which corresponds to the
dimensional counting rules \cite{MMT, BF}. The perturbative QCD
behavior can be preceded by a transition regime $d\sigma/dt \sim
t^{-8}$.

Different models of ``hard" scattering, with power-like scattering
amplitudes obeying scaling, exist in the literature. For example,
in a quark-diquark model of the nucleon \cite{diquarks}, the
introduction of the transverse distance $1/\sqrt{|t|}$ between the
colliding valence quarks gives an additional degree of freedom.

Hard scattering and the power-low behavior of the cross sections
by itself is an important issue of the strong interaction
dynamics, which, however goes beyond the scope of the present
paper. In large-$t$ elastic $pp$ scattering at the LHC one will
enter only the transition region between the ``soft" (exponential)
and ``hard" power-like behavior of the cross section.

   Contrary to Eq.~(\ref{nl-tr}), that predicts the decrease
of the differential cross section in this region, some models lead
to an increasing elastic differential cross section at large
momentum transfer. For example, the DD model \cite{SelyuginPN87_a,
SelyuginPN87_b} describes the elastic differential cross section
not only at small $|t|$, but also up to large $|t| \sim 10 \
$GeV$^2$ at the ISR (see Fig.~\ref{fig:ds14tev.ps}).
Note that the experimental data show that, up to the TeV-region,
the differential cross section has a weak energy dependence at
large $t$. This effect should be better understood, since in the
TeV-region a change of the regime might occur, as
Fig.~\ref{fig:ds14tev.ps} seems to indicate.

\section{Black disc limit at the LHC?}

\bigskip
There is general consensus on the hypothesis that with increasing
energy nucleons are becoming blacker, edgier and larger (BEL
effect). Since the details depend on model calculations and their
extrapolations beyond the energies of present accelerators, the
onset and consequences of BEL effect also differ. Of interest is
the possibility that at the LHC the nucleons will reach the black
disc limit (BDL) with measurable effects. If or when the black
limit will be reached (at the LHC?) two cases, connected with the
use of the $s-$channel unitarity, are possible. According to the
first option, the BDL is identical to the unitarity limit,
absolute for the central opacity of the nucleon. According to the
second option (an alternative solution of the unitarity equation,
which will be illustrated below), the unitarity limit is well
beyond the BDL, leaving space for a further increase with energy
of the central ``overlap function'' (or impact parameter
amplitude). In this approach, the nucleon, after having reached
the BDL and tending to the unitarity limit, will gradually become
more transparent. Below we present both options and their
observable consequences.

\bigskip
\subsection{Definitions}

The onset of ``saturation", wherever it occurs, can hardly be seen
directly from the data, since the behaviour of the real part of the
amplitude is at best known in the nearly forward direction only,
while for the rest it can be only modelled. The expected approach
to the BDL depends both on the model for the scattering amplitude which is
used and on the procedure of unitarization.  Numerical estimates
based on particular model calculations are presented in the two
subsequent subsections.

To start with, let us remind the general definitions and
notations. Unitarity in the impact parameter $b$ representation reads
\begin{equation}
\Im {h(s,b)}=\big|h(s,b)\big|^2+G(s,b). \label{unitarity}
\end{equation}
Here $h(s,b)$ is the elastic scattering amplitude at the center of
mass energy  $\sqrt{s}$,  $\Im h(s,b)$ is usually called the
``profile function'', representing the hadron opacity and $G(s,b)$,
called the ``inelastic overlap function'', is the sum over all
inelastic channel contributions. Integrated over ${b}$, Eq.~(\ref{unitarity})
reduces to a simple relation between total, elastic and
inelastic cross sections:
$$\sigma_{tot}(s)=\sigma_{el}(s)+\sigma_{in}(s).$$
Eq. (\ref{unitarity}) imposes the absolute limit
\begin{equation}
0 \leq \big|h(s,b)\big|^2\leq \Im h(s,b) \leq 1,
\end{equation}
while the so-called black disc limit
$$\sigma_{el}(s)=\sigma_{in}(s)={1\over 2}\sigma_{tot}(s)$$ or
\begin{equation}
\Im h(s,b) = \frac{1}{2}
\end{equation}
is a particular realization of the optical model, namely it
corresponds to the maximal absorption within the eikonal
unitarization, when the scattering amplitude is approximated as
\begin{equation}
h(s,b)={i\over 2}\left(1-\rm{exp}\left[i \ \omega(s,b) \right]\right),
\label{eikonal}
\end{equation}
with a purely imaginary eikonal $\omega(s,b)$.

The solutions of the unitarity equation (\ref{unitarity}),
with $\Re h(s,b)=0$, are
\begin{equation}
h(s,b)={1\over 2}\left[1\pm\sqrt{1-4G(s,b)}\right].
\label{plusminus}
\end{equation}
The solution with the minus sign in front of the square root
describes the eikonal unitarization. The other one, that
with the plus sign in front of the
square root, is known within the so-called $U$-matrix
approach \cite{Khrustalev, tt, TT, VJS}, where the unitarized
amplitude has the form of a ratio rather than of an exponential,
typical of the eikonal approach. It is written as
\begin{equation}
h(s,b)={{U(s,b)}\over{1-i\ U(s,b)}},
\end{equation}
where $U$ is the input Born term, the analogue of the eikonal
$\omega$ in Eq. (\ref{eikonal}). The connection between the two
forms was discussed in recent papers \cite{CSUm, T}.

In the $U$-matrix approach, the scattering amplitude $h(s,b)$ may
exceed the black disc limit as the energy increases. The
transition from a (central) black disc to a (peripheral) black
ring, surrounding a gray disc, for the inelastic overlap function
in the impact parameter space corresponds to the transition from
shadowing to antishadowing \cite{TT}. We shall present a
particular realization of this regime.

The impact parameter amplitude $h(s,b)$ can be calculated either
directly from the data, as it was done e.g. in Ref.~\cite{ASch}
(where, however, the real part of the amplitude was neglected) or
by using a particular model that fits the data sufficiently well.
We consider three representative examples, namely the
Donnachie-Landshoff (DL) model
\cite{Landshofftot_a,Landshofftot_b,DL1_a,DL1_b,DL1_c,DL1_d}, the
DP model \cite{Covolan_a, Covolan_b, VJS} and the DD model
\cite{SelyuginYF80,SelyuginYF82_a,SelyuginYF82_b,SelyuginNP87}.

\bigskip

\subsection{The ``Born term"}

The Donnachie-Landshoff  model
\cite{Landshofftot_a,Landshofftot_b,DL1_a,DL1_b,DL1_c,DL1_d} is
very popular for its simplicity.  Essentially, it means the
following four-parametric empirical formula to fit all total
hadronic cross sections
\begin{equation}
\sigma_{tot}=X\ s^{\delta}+Y\ s^{\delta_r},
\label{empiric}
\end{equation}
where two of the parameters, namely  $\delta=\alpha_P(0)-1\approx
0.08$, and $\delta_r$,
which is negative, are universal. The violation of
the Froissart-Martin (FM) bound,
\begin{equation}
\sigma_{tot}(s) < C\ (\ln {s})^2,\ \ \ \quad C=60\ {\rm mb},
\end{equation}
inherent to this model, is rather an aesthetic than a practical
defect.

The $t$ dependence in the DL model is usually chosen
\cite{DL1_a,DL1_b,DL1_c,DL1_d} in the form close to the dipole
Pomeron form factor. For the present purposes a simple exponential
residue in the Pomeron amplitude will do as well, with the
signature included:
\begin{equation}
A(s,t)= -\ N \left(-i{s\over s_{dl}}\right)^{\alpha(t)} e^{Bt}\ ,
\end{equation}
where $\alpha(t)=\alpha_P(0)+\alpha'\ t $ is the Pomeron
trajectory and $N$ is a dimensionless normalization factor related
to the total cross section at $s=s_{dl}$ by the optical theorem:
\begin{equation}
N = {s_{dl}\over 4\pi\sin{{\pi\over 2}\alpha_P(0)}}\sigma_{tot}(s=
s_{dl})\ .
\end{equation}

\noindent According to fits of Ref.~\cite{DL1_a,DL1_b,DL1_c,DL1_d}
one has $s_{dl}=1$ GeV$^2$, $\alpha_P(0)=1.08$, $\alpha'=0.25$
GeV$^{-2}$, and $X=21.70$ mb (see (Eq.~\ref{empiric})) and thus
$N=4.44$. By puting
\begin{equation}
{d\sigma(s,t)\over dt}\ ={d\sigma(s,t=0)\over dt}\ e^{B_{exp}(s)\ t}
\end{equation}
and choosing the CDF or E410 results for the
slope $B_{exp}$ at the Tevatron energy, we obtain
$B= B_{exp}(s)/2 - \alpha'\ln({{s/s_{dl}}})=4.75$ GeV$^{-2}$.

In the DP model \cite{VJS}, factorizable  at
asymptotically high energies, logarithmically rising cross
sections are produced at the Pomeron intercept equal to one, so
that the DP model does not conflict with the FM bound. While data
on total cross section are compatible with a logarithmic rise
(according to the DP model), the ratio $\sigma_{el}/\sigma_{tot}$
is found (see Ref. \cite{VJS} for details) for $\delta=0$ to be a
monotonically decreasing function of the energy for any physical
value of the parameters. The experimentally observed rise of this
ratio can be achieved only for $\delta > 0$ and thus requires the
introduction of a supercritical Pomeron, with $\alpha(0)>1$. As a
result, the rise of the total cross section is driven and shared
by the dipole Pomeron and the ``supercritical" intercept. The parameter
$\delta=\alpha_P(0)-1$ in the supercritical DP model is nearly
half that of the DL model, making it safer from the point of view
of the unitarity bounds. Generally speaking, the closer the input
to the unitarized output, the better the convergence of the
unitarization procedure.

Let us remind that, apart from the conservative FM bound, any
model should satisfy also $s$-channel unitarity. We demonstrate
below that both the DL and DP models are well below this limit
and will remain so for long time, in particular at LHC. Let us remind
that the DL and the DP model are close numerically, although they
are different conceptually and consequently their extrapolations
to superhigh energies will differ as well.

The elastic scattering amplitude corresponding to the exchange of
a dipole Pomeron reads
\begin{equation}
\begin{array}{rcl}
P(s,t)={d\over{d\alpha}} \left[{\rm
e}^{-i\pi\alpha/2}G(\alpha)(\frac{s}{s_0})^\alpha\right]\\ =& {\rm
e}^{-i\pi\alpha/2}(\frac{s}{s_0})^\alpha
\big[G'(\alpha)+(L-i\pi/2)G(\alpha)\big],
\end{array}
\label{DP1}
\end{equation}
where $L\equiv$ln${\frac{s}{s_0}}$ and
$\alpha\equiv\alpha_P(t)$ is the Pomeron trajectory.

By setting $G'(\alpha)=-a{\rm e}^{b_p(\alpha-1)},$ Eq. (\ref{DP1})
can be rewritten in the geometrical form discussed in Subsection
5.1, (see Eq. (\ref{DP})).

In Table~\ref{table:params_dp} we quote the numerical values of
the parameters of the DP model fitted in Ref.~\cite{Covolan_a,Covolan_b} to the data on
proton-proton and proton-antiproton elastic scattering:
\begin{equation}
\sigma_{tot}(s)={4\pi\over s} \Im A(s,0)\ , \ \rho (s)={\Re
A(s,0)\over{\Im A(s,0)}} \quad ; \qquad 4\;({\rm GeV}) \leq \sqrt s
\leq 1800 \;({\rm GeV})
\end{equation}
as well as the differential cross-section
\begin{equation}
{d\sigma(s,t)\over dt}\ =\ {\pi\over s^2}\left|A(s,t)\right|^2,
23.5\;({\rm GeV})\leq \sqrt s \leq 630\;({\rm GeV}), 0\;({\rm GeV^2}) \leq
|t| \leq 6\;({\rm GeV}^2)\;.
\end{equation}
In that fit, apart from the Pomeron, the Odderon and two
subleading trajectories $\omega$ and $f$ were also included. Here,
for simplicity and clarity we consider only the dominant term at
high energy due to the Pomeron exchange  with the parameters
fitted in Ref.~\cite{Covolan_a,Covolan_b}.

\begin{table}

\medskip
\begin{center}
\begin{tabular}{|c|c|c|c|c|c|}
\hline $a$ & $b_p$ & $\alpha_P(0)$ & $\alpha'$(GeV$^{-2}$)  &
$\epsilon$ &
$s_0$(GeV$^2$)\\
\hline
355.6 & 10.76 & 1.0356 & 0.377 & 0.0109 & 100.0\\
\hline
\end{tabular}
\end{center}
\caption{Values of the parameters of the DP model, Subsections 5.1
and 7.2, found in
Ref.~\cite{Covolan_a,Covolan_b}.\label{table:params_dp}}
\end{table}

We use the above set of parameters to calculate the impact
parameter amplitude.

\bigskip

\subsection{Impact parameter representation, the black disc
limit\\ and unitarity}

The elastic amplitude in the impact parameter representation in
our normalization is
\begin{equation}
h(s,b)\ = \ {1\over 2 s} \int^\infty _0 dq\ q J_0(b q) A(s,-q^2), ~~~
~\quad  \ q=\sqrt{-t}\ , \label{Bessel}
\end{equation}
expressed in tems of the Bessel function $J_0$.

The impact parameter representation for linear trajectories
\footnote {Similar calculations with non-linear trajectories can
be found e.g. in Refs.~\cite{VJS, JP_NC}.} is calculable
explicitly for the DP model using Eq.~(\ref{DP1}). We have
\begin{equation}
h(s,b)\ =i\ g_0\ \big[ e^{r_1^2\delta} \ e^{-{b^2 /4 R_1^2}}\ -\
\epsilon\ e^{r_2^2\delta} \ e^{-{b^2 / 4 R_2^2}}\big]\ ,
\end{equation}
where the functions $r_i^2\equiv r_i^2(s) (i=1,2)$ have
been introduced in the Subsection 5.1 and
\begin{equation}
R_i^2= \alpha' r_i^2~,\ \ \ g_0= {a\over 4 b_p\alpha' s_0}\ .
\end{equation}
Asymptotically (i.e. when $L\gg b_p$, which implies $\sqrt s \gg 2. $ TeV,
with the parameters of Table~\ref{table:params_dp}) we get
\begin{equation}
h(s,b)\rightarrow i\ g(s)\ (1-\epsilon)\ e^{-{b^2\over 4 R^2}}, \
\ s\to\infty,
\end{equation}
where
\begin{equation}
R^2=\alpha'L~,\ \ \ \  g(s)=g_0\left({s\over
s_0}\right)^\delta \ .
\end{equation}

To illustrate the effect coming from $s$-channel unitarity, in
Ref. \cite{JenkovszkyYF00, DJStr} a family of curves showing the
imaginary part of the amplitude in the impact
parameter-representation as well as the inelastic overlap function
$G(s,b)$ calculated from Eq.~(\ref{unitarity}) at various
energies, was displayed.

Our confidence in the extrapolation of $\Im h(s,b)$ to the highest
energies rests partly on the good agreement of our (non fitted)
results with the experimental analysis of the central opacity of
the nucleon  (see Table~\ref{table:opacity}).

\begin{table}

\medskip
\begin{center}
\begin{tabular}{|c|c|c|c|} \hline
$\sqrt{s}$ & 53 GeV & 546 GeV  & 1800 GeV \\
\hline
exp& 0.36 & $0.420\pm 0.004$& $0.492\pm 0.008$ \\
th& 0.36 & 0.424 & 0.461 \\
\hline
\end{tabular}
\end{center}
\caption{Central opacity of the nucleon $\Im h(s,0)$ calculated at
ISR, SPS and Tevatron energies compared with experimental results.
\label{table:opacity}}
\end{table}

It is important to note that the unitarity bound $1$ for $\Im h(s,b)$
will not be reached at the LHC energy, while the black disc limit
$1/2$ will be slightly exceeded, the central opacity of the nucleon
being $\Im{\rm m}h(s,0)=0.54$.

The black disc limit is reached at $\sqrt{s}\sim 2 $ TeV, where
the overlap function reaches its maximum value $1/4$. This energy
corresponds to the appearance of the antishadow mode in agreement
with the general considerations in Ref.~\cite{TT}. Notice that while
$\Im h(s,b)$ remains central all the way, $G(s,b)$ is getting
more peripheral as the energy increases starting from the the value of
Tevatron. For example at $\sqrt{s}=14$ TeV, the central region of
the antishadowing mode, obtained from the $U$ matrix
unitarization, below $b\sim 0.4$ fm is discernible from the
peripheral region of shadowing scattering beyond $b\sim 0.4$ fm,
where $G(s,b)={1/4}$. Consistently with the BEL effect the proton will
tend to become more transparent at the center (in the sense
of becoming a gray object surrounded by a black ring), i.e. it is
expected to become gray, edgier and larger (GEL effect).

The $s$ channel unitarity limit will not be endangered until
extremely high energies ($10^5$ Gev for the DL model and $10^6$ GeV
for the DP model), safe for any credible experiment. It is interesting
to compare these limits with the one imposed by the
FM bound:  actually the Pomeron amplitude saturates
the FM bound at $10^{27}$ GeV. As expected, the FM bound is even
more conservative than that following from $s$-channel unitarity.

Now, we consider the unitarized amplitude according to the
$U$-matrix prescription \cite{Khrustalev, tt, VJS}
\begin{equation}
H(s,b)={h(s,b)\over{1-ih(s,b)}}~,
\end{equation}
with the Born term $h(s,b)$ calculated from Eq.~(\ref{Bessel}) using
Eq.~(\ref{DP1}).

An unescapable consequence of the unitarization is that, when
calculating the observables, one should also replace the Born
amplitude $A(s,t)$ with a unitarized amplitude $\widetilde {A}(s,t)$
defined as the inverse Fourier-Bessel transform of $h(s,b)$:
\begin{equation}
\widetilde {A} (s,t)\ = \ {2 s} \int^\infty _0 db\ b J_0(b
\sqrt{-t}) H(s,b)\ .
\end{equation}
Thus, the above picture may change since the parameters of the
model should in principle be refitted under the unitarization
procedure (this effect of changing the parameters was clearly
demonstrated e.g. in Ref.~\cite{CGM}).

While the unitarity limit now is secured automatically (remind
that $\Im h(s,0)$ is well below that limit even at the Born level
in the TeV region), the behaviour of the elastic impact parameter
amplitude after it has reached the black disc limit corresponds
(see Ref.~\cite{TT}) to the transition from shadowing to antishadowing.
In other words, the proton (antiproton) after having reached its
maximal blackness around $2$ TeV, will become gradually more
transparent with increasing energies at its center. It follows
from the presented model that, in getting edgier and
larger, the proton, after reaching its maximal blackness, will
tend to be more transparent (GEL effect), i.e. a gray disc surrounded by
a black ring will gradually develop beyond the Tevatron energy
range. The transition from shadowing to a new, antishadowing
scattering mode is expected to occur at the LHC.

To conclude, we stress once more that both the data and relevant
models at present energies are well below the $s$-channel
unitarity limit. Deviations due to the diversity of realistic
models may result in discrepancies concerning $\Im{h(s,0)}$ of at
most $10\%$, while its value at 2 TeV is still half that of the
unitarity limit, so there is no reason to worry about it. Opposite
statements may result from the confusion with normalization.
Therefore, the model amplitudes at the Born level may still be quite
interesting and efficient in analyzing the data at energies of present
accelerator and giving some predictions beyond. The
question, which model is closer to reality and meets better the
requirements of the fundamental theory remains of course topical.

\subsection{Saturation at the LHC (in the DD model)}

The effect of the Black Disc Limit (BDL) saturation can be
observed in the change of the $t$-dependence of the slope $B$ and
of the ratio $\rho$ beginning from approximately $\sqrt{s} = 2
\div 6 \ $TeV and tending to enhance at  $\sqrt{s} = 14 \ $TeV
\cite{SelyuginBDL06}. Such a saturation effect was studied in the
framework of the DD model.

In Refs.~\cite{SelyuginBDL06,SelyuginBDLCJ04} the BDL is
determined by account of a hard Pomeron. The model predicts the increase
with $|t|$ of the slope $B(t)$ at small $t$ and energies typical of the LHC, as
shown in Fig.~\ref{fig:rosc.ps}, right panel.
In the same Figure, left panel, it is shown that saturation of the BDL
substantially modifies the $t$-dependence of the ratio $\rho(s,t)$ (see,
also, Table~\ref{table:saturation}).

\begin{table*}
\begin{tabular}{llll}
\hline\noalign{\smallskip}
\multicolumn{4}{c} {${\rho}  \ (\sqrt{s} =14 \ $TeV, t)}   \\
\noalign{\smallskip}\hline\noalign{\smallskip}
\multicolumn{2}{c} {$ t= 0 \ $  GeV $^2$} &  \multicolumn{2}{c} {$ t= - 0.1 \ $ GeV $^2$} \\

{DD  \cite{SelyuginPN87_a,SelyuginPN87_b}} & Soft and hard Pomeron \cite{SelyuginBDL06} & DD  & Soft and hard Pomeron \\
  $0.19 $     & $0.24 $      & $0.08 $ & 0.05 \\
\noalign{\smallskip}\hline
\end{tabular}
\caption{The effect of saturation on the ratio $\rho(t)$ as
predicted in Refs. \cite{SelyuginPN87_a,SelyuginPN87_b},
\cite{SelyuginBDL06}.\label{table:saturation}}
\end{table*}

\begin{figure}[tbh!]
\includegraphics[width=0.5\textwidth]{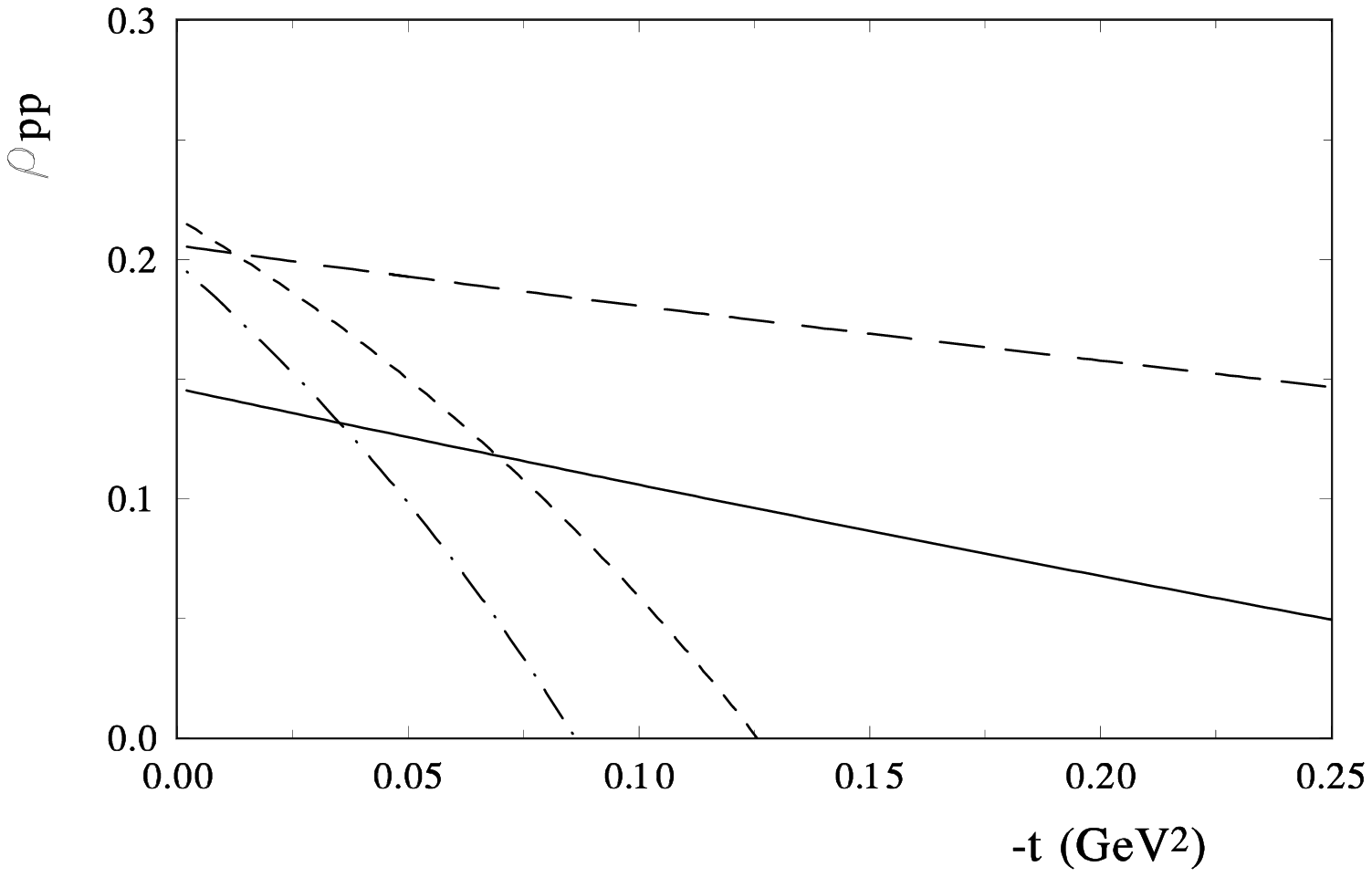}
\includegraphics[width=0.5\textwidth]{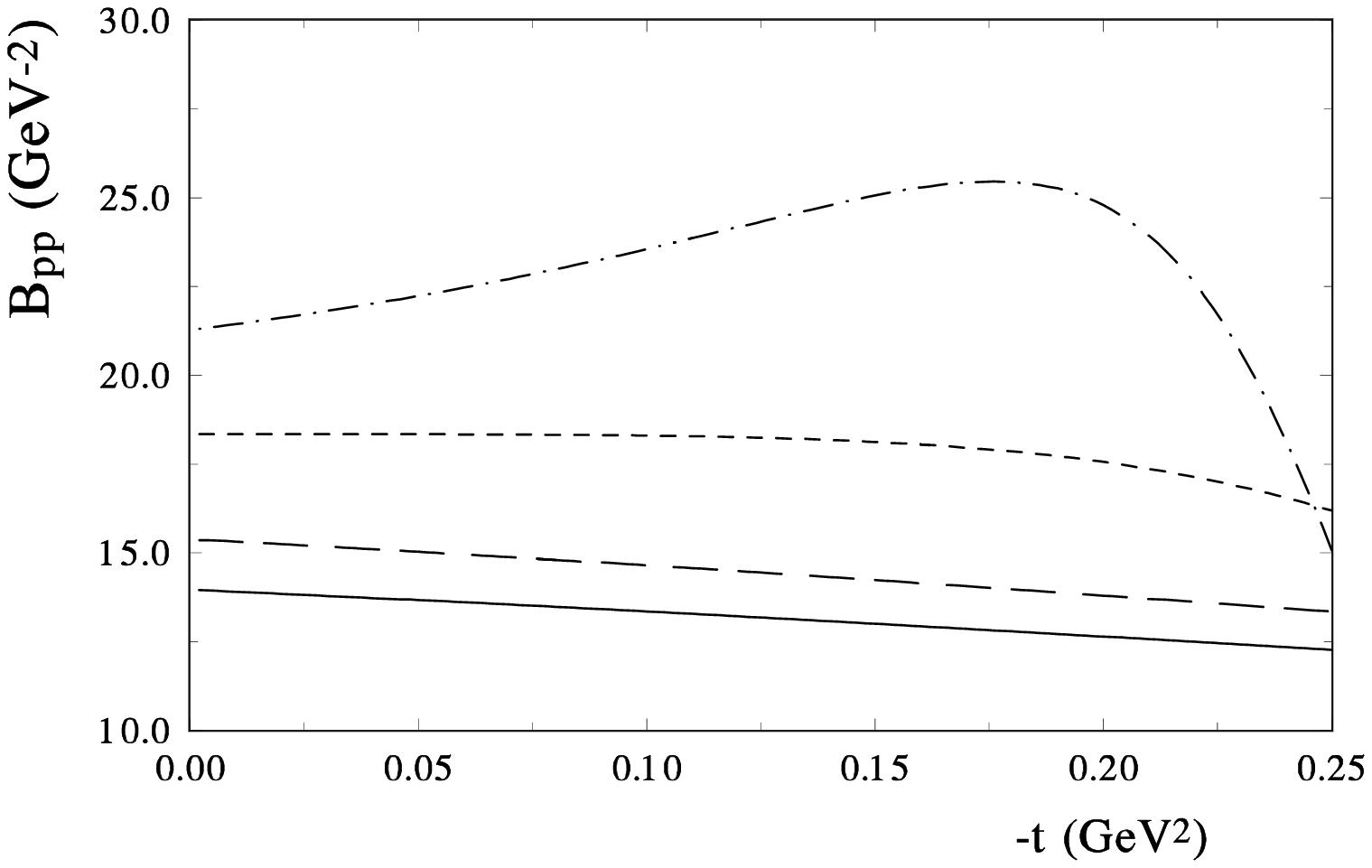}
 \caption{{Left
panel}: The ratio of the real to the imaginary part of the
amplitude as a function of $t$, for the bare and the saturated
amplitudes at various energies: 100~GeV (solid line), 500~GeV
(long dashes), 5~TeV (short dashes) and 14~TeV (dash-dotted
curve). {Right panel}: The slope of the elastic differential
cross section as a function of $t$, for the bare and saturated
amplitudes at various energies: $100$ GeV (solid line), $500$ GeV
(long dashes), 5 TeV (short dashes) and 14 TeV (dash-dotted
curve).} \vspace{0.5cm} \label{fig:rosc.ps}
\end{figure}

In fact, saturation naturally predicts a small increase of the
slope $B(t)$ with $t$ at small $t$. To understand this phenomenon, let us
take a simple black disk form with a sharp edge at radius $R$. The
relevant scattering amplitude can then be written as
\begin{eqnarray}
    A(s, t = 0) \ \sim \ \frac{J_{1}(\sqrt{-t} R)}{\sqrt{-t} R} \ .
    \label{hbd}  \nonumber
\end{eqnarray}

In this model the slope of the differential cross section at small
momentum transfer behaves as

\begin{equation}
  B_{BDL} \sim \frac{R^2}{4}  + \left(\frac{R^4}{32}\right) |t|.  \label{b-bdl}
\end{equation}

Hence the slope grows with increasing $|t|$ at small momentum transfer,
as one can see in Fig.~\ref{fig:rosc.ps}.
    It is interesting that two quite different models lead practically to the same
  result.

Saturation is usually attibuted to the growing gluon density at
small $x$. This phenomeonon is important at relatively large
impact parameters, where the role of non-perturbative effects
\cite{kovner-cf}, due to confinement, is not clear. On the other
hand, the BDL-saturation is generally connected with the small
impact parameter region. Scrutinizing the small $|t|$ region and
looking for possible saturation effects there \cite{BDLLHC} is of
great interest for future LHC measurements.

\section{Summary and outlooks} \label{s6}

In the present paper we concentrated mainly on elastic scattering and
total cross sections.
It should be remembered that the
elastic scattering amplitude is the central, in a sense, basic
object of the theory, therefore its knowledge is essential for the
further progress in understanding strong interactions.
Furthermore, elastic scattering, by unitarity (see, e.g. Sec. 7),
is directly connected to inelastic processes.

The LHC facilities provide a unique experimental environment where
one can test properties of the Pomeranchuk trajectory. Let us
remind the open questions and problems that can be addressed both
theoretically and experimentally:

$\bullet$ The Pomeron intercept. From perturbative QCD the value
of about $\alpha(0)\approx 1.25$ follows
\cite{BFKL_a,BFKL_b,BFKL_c,BFKL_d}. It should be remembered, as
often stressed by the authors of the BFKL equation, that this
value should not be confronted with the data prior the
unitarization of the amplitude (that will lower the rate of
increase of the total cross section). A complete unitarization
procedure, however, is difficult to be realized. In any case,
experimental measurement of the total cross section will shed new
light on the value of the Pomeron intercept.

$\bullet$ Similar arguments concern the slope of the trajectory,
which, because of its non-linear nature, is $t$-dependent. Hence
it is important to explore the small $|t|$ region of the elastic
differential cross section.

$\bullet$ By parameterizing the diffraction cone by the method of
overlapping bins, the fine structure of the cone (or that of
the Pomeron) will be detected (Sec. 3).

$\bullet$ Perturbative QCD predicts logarithmic asymptotics for
the Pomeron trajectory, as in Eq.~(3). This basic result of QCD is
often ignored. At the LHC, the onset of, or approach to the
logarithmic behavior of the Pomeron trajectory should be visible.

$\bullet$ The ``QCD Pomeron" \cite{BFKL_a,BFKL_b,BFKL_c,BFKL_d}
implies an infinite number of poles or a moving cut, which are
difficult to be realized phenomenologically. In a simplified
version, they can be approximated as a sum of two poles with
different intercepts.

$\bullet$ The shape and the depth of the expected diffractive
minimum and its ratio to the height of the second maximum are
extremely informative (Sec. 5). The theoretical predictions here
are not unique, but future data from LHC will narrow the margin
for existing theoretical models.

$\bullet$ The expected transition from soft to hard physics will
be revealed at the LHC. The determination of the $s$ dependence of
the elastic cross section for $|t|>1$ GeV$^2$ will be very
important for understanding the onset of the  new, ``hard"
dynamics, replacing the ``soft", diffractive one.

$\bullet$ A basic characteristics of the strong interaction
dynamics is the ratio $R(s)=\sigma_{el}/\sigma_t$, reflecting the
intensity of particle production and/or the realization of a
particular unitarization scheme. In the past, $R(s)$ was changing
unpredictably: from the (seemingly asymptotic) constant in the ISR
energy region, suggesting geometrical scaling, to a subsequent
indefinite rise. Will this trend continue or a "new asymptotic"
plateau is expected at the LHC?

\medskip

\section{Acknowledgements}
We thank Mario Deile, Alan Martin, Marcio Jose Menon, Francesco
Paccanoni, Vladimir Petrov, Jacques Soffer and Serguey Troshin for
useful remarks. L.J. and O.S. are grateful to the Department of
Theoretical Physics of the University of Torino and the Department
of Physics of the University of Calabria and the Istituto
Nazionale di Fisica Nucleare - Sezione di Torino and Gruppo
Collegato di Cosenza, where part of this work was done, for their
warm hospitality and support. R.O. wants to express his gratitude
to the Academy of Finland for support. A.P. and R.F. acknowledge
partial support by the Italian Ministry of University and
Research.

\vfill \eject
\end{document}